\documentclass[a4papers,runningheads]{llncs}

\usepackage{hyperref}
\usepackage[firstpage]{draftwatermark}



\newif\ifblind
\blindfalse

\newif\ifdraft
\draftfalse

\newif\iflong
\longfalse

\newif\ifarxiv
\arxivtrue

\usepackage[T1]{fontenc}
\usepackage{microtype}
\usepackage{newtxtext,newtxmath}
\usepackage[scaled=0.81]{beramono}

\usepackage{graphicx}
\usepackage{xcolor,xstring}

\usepackage{tikz}
\usetikzlibrary{shapes,arrows,arrows.meta,chains,positioning,calc,fit,intersections,through}

\definecolor{rangercol}{HTML}{008837}
\definecolor{commentcol}{RGB}{60,60,60}

\usepackage[inline]{enumitem}

\usepackage{ragged2e}
\usepackage{booktabs}
\usepackage{relsize}
\usepackage{changepage}            %
\usepackage[skip=4pt,font=small,strut=off]{caption}
\usepackage[skip=2pt]{subcaption}  %

\newcommand{\captionskip}[1]{
    \caption{#1}
    \vspace{2mm}
}

\usepackage{listings}

\lstdefinestyle{displayed}{
  numbers=left, %
  firstnumber=last, 
  breaklines=true,
  numbersep=2pt,
  frame=tb,
  numberstyle=\tiny,
  tabsize=2,
  captionpos=b,
  xleftmargin=0mm, %
  xrightmargin=0mm, %
  basicstyle=\ttfamily\scriptsize,
  keywordstyle=\bfseries\ttfamily,
  keepspaces=true,
  columns=fixed,
  mathescape=true,
  showstringspaces=false
}

\lstdefinestyle{compact}{
  numbers=none, %
  frame=none,
  xleftmargin=0mm, %
  xrightmargin=0mm, %
  basicstyle=\ttfamily\scriptsize,
  keywordstyle=\bfseries\ttfamily,
  keepspaces=true,
  columns=fixed,
  mathescape=true,
  showstringspaces=false
}

\newcommand{\thickdelim}[1]{\scalebox{1.9}[1]{#1}}

\newcommand{\inclif}[2]{\ensuremath{\{#1 \mid #2\}}}

\newcommand{\rulen}[1]{{\smaller\textsf{#1}}}
\newcommand{\pgroup}[1]{{\smaller\textsf{#1}}}
\newcommand{\pex}[1]{\pgroup{#1}}

\lstdefinelanguage{Licorne}{%
  morekeywords=[1]{%
as,class,datatype,else,fn,for,if,import,interface,is,main,new,null,object,package,panic,private,public,pure,record,return,sub,super,then,this,typealias,val,var,when,where,while,true,false,null %
  },%
  morekeywords=[2]{Int,Double,Char,Bool,String,Any,Nothing,Unit},   %
  morekeywords=[3]{NonZeroLength},   %
  morekeywords=[4]{it,with},                 %
  literate={%
    {->}{{$\rightarrow$}}{1}
    {!!}{{$\text{!}\!\text{!}$}}{1}
    {<<}{{$\color{rangercol}\thickdelim{[}$}}{1}
    {>>}{{$\color{rangercol}\thickdelim{]}$}}{1}
    {..}{{\textcolor{rangercol}{\textbf{..}}}}{3}
    {..<}{{\textcolor{rangercol}{\textbf{..<}}}}{4}
  },
  basicstyle={\ttfamily\scriptsize},
  keywordstyle={[1]\bfseries},
  keywordstyle={[2]}, %
  keywordstyle={[3]\color{rangercol}},
  keywordstyle={[4]\color{rangercol}\bfseries},
  keepspaces,
  morecomment=[l]{//},
  morecomment=[s]{/*}{*/},
  commentstyle={\color{commentcol}\itshape},
  escapeinside={(@}{@)}
}[keywords,comments,strings]%

\lstdefinelanguage{IRcorne}{%
  morekeywords=[1]{%
    if,then,else,loop,pure,fn,return,panic,call,read,write,is,as,true,false,unit,merge,next,while,do,from
  },%
  morekeywords=[2]{Int,Double,Char,Bool,String,Any,Nothing,Unit},
  morekeywords=[3]{it},
  morekeywords=[4]{with},
  literate={%
    {->}{{$\rightarrow$}}{1}
    {!!}{{$\text{!}\!\text{!}$}}{1}
    {<<}{{$\color{rangercol}\thickdelim{[}$}}{1}
    {>>}{{$\color{rangercol}\thickdelim{]}$}}{1}
    {..}{{\textcolor{rangercol}{\textbf{..}}}}{2}
    {..<}{{\textcolor{rangercol}{\textbf{..<}}}}{2}
  },
  basicstyle={\ttfamily\scriptsize},
  keywordstyle={[1]\color{commentcol}\bfseries},
  keywordstyle={[2]\color{commentcol}}, %
  keywordstyle={[3]\color{rangercol}},
  keywordstyle={[4]\color{rangercol}\bfseries},
  keepspaces,
  morecomment=[l]{//},
  morecomment=[s]{/*}{*/},
  commentstyle={\color{commentcol}\itshape},
  escapeinside={(@}{@)}
}[keywords,comments,strings]%

\newcommand{\Lic}[1]{\mbox{\lstinline[language=Licorne,basicstyle=\ttfamily\smaller]|#1|}}

\newcommand{\J}[1]{\mbox{\lstinline[language=Java,basicstyle=\ttfamily\smaller]|#1|}}

\newcommand{\IRc}[1]{\mbox{\lstinline[language=IRcorne,basicstyle=\ttfamily\smaller]|#1|}}

\DeclareDocumentCommand{\IRKW}{ m m }{%
  \IfEqCase{#1}{
    {0}{\ensuremath{\texttt{#2}}}
    {1}{\ensuremath{\texttt{\color{commentcol}\bfseries #2}}}
    {2}{\ensuremath{\texttt{#2}}}
    {3}{\ensuremath{\texttt{\color{rangercol}#2}}}
    {4}{\ensuremath{\texttt{\color{rangercol}\bfseries #2}}}
    }[#2]%
}

\newcommand{\IRas}{\IRKW{1}{as}}
\newcommand{\IRis}{\IRKW{1}{is}}
\newcommand{\IRif}{\IRKW{1}{if}}
\newcommand{\IRthen}{\IRKW{1}{then}}
\newcommand{\IRelse}{\IRKW{1}{else}}

\newcommand{\IRfn}{\IRKW{1}{fn}}
\newcommand{\IRreturn}{\IRKW{1}{return}}
\newcommand{\IRpanic}{\IRKW{1}{panic}}
\newcommand{\IRcall}{\IRKW{1}{call}}

\newcommand{\IRmerge}{\IRKW{1}{merge}}
\newcommand{\IRnext}{\IRKW{1}{next}}
\newcommand{\IRfrom}{\IRKW{1}{from}}
\newcommand{\IRwhile}{\IRKW{1}{while}}
\newcommand{\IRloop}{\IRKW{1}{loop}}
\newcommand{\IRInt}{\IRKW{2}{Int}}
\newcommand{\IRBool}{\IRKW{2}{Bool}}

\newcommand{\IRNothing}{\IRKW{2}{Nothing}}
\newcommand{\IRUnit}{\IRKW{2}{Unit}}

\newcommand{\IRbb}{\IRKW{2}{!\!!}}
\newcommand{\IRit}{\IRKW{3}{it}}
\newcommand{\IRwith}{\IRKW{4}{with}}

\newcommand{\IRlc}{\IRKW{3}{\thickdelim{[}}}

\newcommand{\IRrc}{\IRKW{3}{\thickdelim{]}}}
\newcommand{\IRdd}{\IRKW{3}{..}}

\newcommand{\IR}[1]{\IRKW{0}{#1}}

\newcommand{\subtype}{\ensuremath{\mathrel{<:}}}
\newcommand{\returns}{\ensuremath{\mathrel{\Uparrow}}}
\newcommand{\uniontype}{\ensuremath{\mathrel{\sqcup}}}
\newcommand{\intertype}{\ensuremath{\mathrel{\sqcap}}}

\newcommand{\rep}[2]{\ensuremath{\left\langle#1\right\rangle^{#2}}}

\newcommand{\cand}{\ensuremath{\textsc{cnd}}}
\newcommand{\cout}[2][\cand]{\ensuremath{#1_{#2, \mathsf{out}}}}
\newcommand{\cin}[2][\cand]{\ensuremath{#1_{#2, \mathsf{in}}}}
\newcommand{\cplain}[2][\cand]{\ensuremath{#1_{#2}}}

\usepackage{amsmath,stmaryrd}
\usepackage{bussproofs}

\usepackage{mathpartir}

\usepackage{tablefootnote}

\usepackage{multicol,multirow}
\usepackage{xurl,xspace}

\ifarxiv
  \usepackage{fontawesome5}
  \newcommand{\faCircleCheck}{\faCheckCircle}
  \newcommand{\faXmark}{\faTimes}
\else
  \usepackage{fontawesome7}
\fi

\newcommand{\failure}{\ensuremath{-}}

\usepackage{makecell}

\iflong
\else
\makeatletter
\newcommand\nicepar{\@startsection{paragraph}{4}{\z@}%
  {-1\p@ \@plus -1\p@ \@minus -1\p@}%
  {-0.3em \@plus -0.1em \@minus -0.1em}%
  {\itshape\normalsize}}
\newcommand\nicesection{\@startsection{section}{1}{\z@}%
  {-8\p@ \@plus -3\p@ \@minus -3\p@}
  {4\p@ \@plus 2\p@ \@minus 1\p@}
  {\normalfont\large\bfseries\boldmath}}
\newcommand\nicesubsection{\@startsection{subsection}{2}{\z@}%
  {-5\p@ \@plus -2\p@ \@minus -2\p@}%
  {2\p@ \@plus 1\p@ \@minus 1\p@}%
  {\normalfont\normalsize\bfseries\boldmath}}
\newcommand\nicesubsubsection{\@startsection{subsubsection}{3}{\z@}%
  {-3\p@ \@plus -1\p@ \@minus -1\p@}%
  {-0.3em \@plus -0.1em \@minus -0.1em}%
  {\normalfont\normalsize\bfseries\boldmath}}
\makeatother
\renewcommand{\subsubsection}{\nicesubsubsection}
\renewcommand{\subsection}{\nicesubsection}
\renewcommand{\section}{\nicesection}
\renewcommand{\paragraph}{\nicepar}
\fi

\ifdraft
\usepackage[colorinlistoftodos,textsize=scriptsize,obeyFinal,textwidth=33mm]{todonotes}

\DeclareDocumentCommand{\ReviewNote}{s o m O{white}}{%
  \todo[color=#4,\IfBooleanTF{#1}{inline}{}]{\IfNoValueF{#2}{\textbf{#2:}\xspace}#3}
}
\else
\DeclareDocumentCommand{\ReviewNote}{s o m O{white}}{}
\fi

\DeclareDocumentCommand{\caf}{s m}{\IfBooleanTF{#1}{\ReviewNote*{#2}[yellow]}{\ReviewNote{#2}[yellow]}}
\DeclareDocumentCommand{\va}{s m}{\IfBooleanTF{#1}{\ReviewNote*{#2}[red!70!white]}{\ReviewNote{#2}[red!70!white]}}

\newcommand{\licorne}{{\smaller[0.5]{\textsc{li\-corne}}}\xspace}
\newcommand{\ranger}{{\smaller[0.5]{\textsc{rang\-er}}}\xspace}
\newcommand{\ircorne}{{\smaller[0.5]{\textsc{IR\-corne}}}\xspace}

\begin{document}

\SetWatermarkText{\hspace*{3.7in}\raisebox{8.55in}{\href{https://eapls.org/eapls/artifact-badges/}{\mbox{\includegraphics[width=1.4cm]{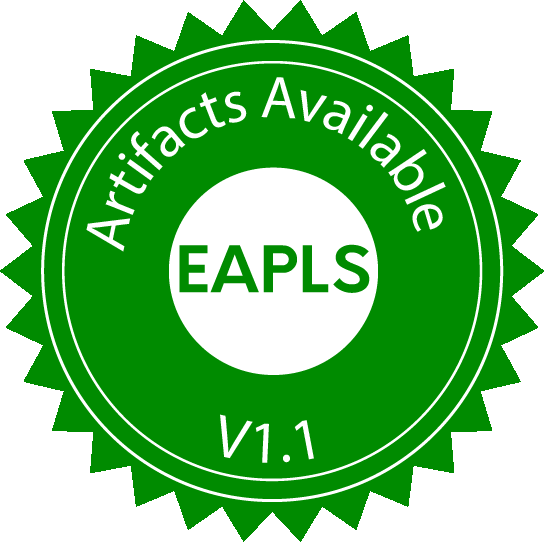}\includegraphics[width=1.4cm]{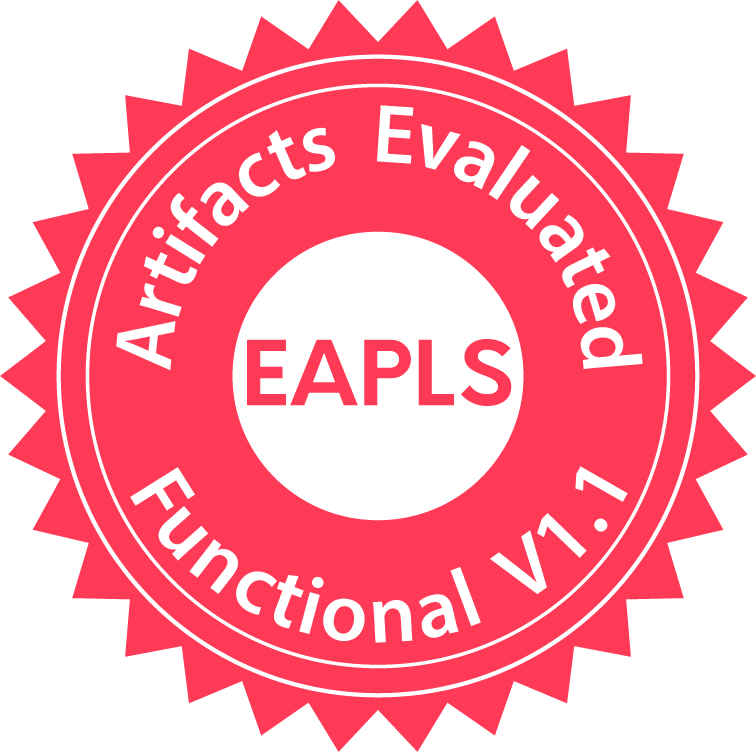}}}}}
\SetWatermarkAngle{0}

\title{Language-level Support for Refinement Types and Range Types Inference}
\title{Practical Range Refinement Types with Inference}
\author{Valentin Aebi\orcidID{0009-0000-7981-1743} \and
Carlo A. Furia\orcidID{0000-0003-1040-3201}}
\authorrunning{V. Aebi and C. A. Furia}
\institute{Software Institute, USI -- Università della Svizzera italiana, Lugano, Switzerland\\
\email{\{valentin.aebi,furiac\}@usi.ch}}
\maketitle              %
\begin{abstract}
  Refinement types are a static verification technique
  that aims at increasing the expressivity of
  traditional type systems while remaining easy and natural to use.
  While systems based on refinement types have been developed
  for several mainstream languages, their practical
  adoption remains limited by their annotation overhead,
  which is often a more significant burden than
  when using the ``plain'' type annotations
  of languages like Java or Scala.

  To improve the state of the art,
  this paper introduces \ranger: a refinement type system
  designed to keep the annotation overhead small and to seamlessly integrate with 
  imperative-style constructs like variables and loops. 
  As the name suggests, \ranger focuses on integer \emph{range types}:
  a particular kind of refinement types that express bounded integer ranges.
  Such types are widely useful to verify correct index manipulation
  and in-bounds data accesses, among others.
  To combine expressiveness and succinctness,
  \ranger is based on a \emph{bidirectional} type system,
  which runs a type inference algorithm
  to provide the typechecking pass 
  with information useful to reduce the need for user-written
  auxiliary annotations.
  \ranger also integrates other forms of lightweight
  flow-sensitive static analysis techniques
  that precisely capture the program's behavior
  without explicit annotations.
 
  We implemented \ranger on top of the \licorne experimental programming language.
  Our experiments show that \ranger's implementation
  can concisely express and verify
  a variety of useful properties that fall beyond the capabilities of standard static type systems like those of Java and Scala.
  The experiments also indicate
  that \ranger compares favorably to
  other extended type systems,
  such as the Java Checker Framework and Liquid Java,
  that can also check properties about ranges:
  \ranger can express more complex constraints,
  and usually requires no auxiliary annotations
  beyond those to specify a function's typed signature.

\end{abstract}

\section{Introduction}

Static type systems
are an integral part
of many programming languages,
where they systematically rule out, at compile time,
the presence of common programming errors.
As such, they have earned a reputation
as ``the world's most successful formal method''~\cite{JonesWEV16}.
Key to their success is a delicate balance
between expressiveness and ease of use: %
the static type systems most commonly found in mainstream programming languages
target a relatively narrow class of safety properties%
---such as operation compatibility and syntactic subtyping---%
that can be expressed and analyzed
with only a reasonable amount of explicit user annotations.
In contrast,
statically verifying more complex, behavioral properties
usually requires specialized tools and detailed annotations, 
which makes them impractical for casual or non-expert users.

Refinement types are an attempt to push the expressiveness
of static type systems while retaining their ease of use.
A refinement type uses a \emph{predicate} to further constrain the 
set of values described by a type.
In this paper, we focus on a category of refinement types
that is widely useful: \emph{integer range types}, 
which constrain integer values to be within certain bounds. 
For example, an array index variable should be 
between zero (included) and the array's length (excluded).
As we discuss in \autoref{sec:related-work},
several frameworks have been developed in recent years
to extend mainstream programming languages
with support for refinement types.
For example, the Checker Framework~\cite{java_checker_framework}
and Liquid Java~\cite{liquid_java}
offer annotations to refine Java types,
both supporting at least some form of ranges. 
Our experiments described in \autoref{sec:evaluation}, however, 
indicate that existing systems may
not ``play well'' with some features of the language,
require a significant number of additional user-written annotations,
or have limited expressivity%
---so that expressing the ``right'' range type constraints
is sometimes cumbersome or impossible.

In this paper,
we introduce \ranger, 
a novel approach for range types
with high precision and limited annotation overhead. %
The design of \ranger is based on two key principles:
\begin{enumerate*}[label=\emph{\roman*})]
\item
  annotations should cover common programming idioms naturally;
  and
\item they should be succinct,
  requiring as little overhead as possible compared to ``regular'' type
  annotations.
\end{enumerate*}
To this end, \ranger deploys complementary techniques:
\begin{enumerate*}[label=\emph{\roman*})]
\item \emph{bidirectional} typechecking~\cite{bidirectional_typing}
  combines top-down type inference and bottom-up type-checking
  to reduce the amount of required annotations;
\item a \emph{monotonicity} analysis improves the precision
  of type inference in the presence of loops;
\item \emph{smart casts} enhance expressiveness 
  by taking flow-sensitive\footnote{
    Typing rules for smart casts are actually \emph{path sensitive};
    however, since the typing literature usually simply designates
    such rules as ``flow-sensitive''~\cite{Callauetal2014},
    we stick to this less precise terminology.
  } information into account;
\item a \emph{hybrid cast} operator
  provides a runtime escape hatch in the few cases when the refinement types
  cannot be verified statically, which provides additional flexibility.
\end{enumerate*}

We implemented the \ranger technique as part of the type system of the \licorne 
experimental programming language: a heavily redesigned variant
of a prototype language we introduced in previous 
work~\cite{valentin_msc_thesis}.
Using \licorne helped to quickly prototype our type system design
and to integrate refinements as seamlessly as possible into 
the type system;
however, the design of \ranger is general and independent of \licorne,
and could be implemented also in other programming frameworks
with similar features.

\autoref{sec:evaluation} describes an experimental evaluation
where we compared our implementation of \ranger to 
other refinement type frameworks that support similar range annotations, 
the Checker Framework and Liquid Java, as well as to the type system of Scala.
These experiments indicate that \ranger
is generally more succinct (requires fewer user annotations)
and/or more expressive (can specify and check more complex constraints)
than the other frameworks.
These results suggest that \ranger achieves
a practical trade-off between expressiveness and ease of use.

\paragraph{Contributions.}
This paper makes the following contributions:
\begin{itemize}
\item \ranger: a practical type system for range refinement types that
  combines expressiveness and succinctness thanks to advanced type inference;
\item an implementation of \ranger, and an experimental comparison
  with other similar range refinement type frameworks;
\item for reproducibility, the prototype implementation of \ranger
  and all examples used in the experiments are available
  in a replication package:\\ \url{https://doi.org/10.6084/m9.figshare.33022760}.
\end{itemize}

\section{Related Work}
\label{sec:related-work}

\subsubsection{Dependent types.}
Type systems are pervasive in programming and, more generally, computer science~\cite{Pierce2002}. 
While most programmers are familiar with
``vanilla'' static type systems as they are used in
mainstream programming languages,
the full power of type systems goes well beyond distinguishing 
floating-point numbers from integers at compile-time.
In fact, so-called \emph{dependent} types%
---whose definition depends on program terms, such as a predicate in the case of refinement types---%
have been introduced as a way of formalizing logic reasoning
that is amenable to automated reasoning~\cite{Bruijn70,intuitionistic},
and underlie powerful interactive theorem provers
such as Coq/Rocq~\cite{coq} and Idris~\cite{idris}.

\subsubsection{Refinement types.}
More recently, there has been a growing interest in applying refinement types
to general-purpose programming languages while retaining ease of use. In particular, 
\emph{liquid types} are refinement types whose predicates are restricted to
a decidable logic fragment.
Liquid types have been implemented on top of
functional programming languages such as 
ML~\cite{liquid_types_ml}, Ocaml~\cite{liquid_types_ocaml}, and Haskell~\cite{liquid_haskell}%
---the latter arguably the most developed implementation of liquid types
for a real-world programming language. Still in the realm of functional languages, 
Aeon~\cite{aeon} has supported refinement types from the very beginning of its development, 
focusing on their use for program synthesis.

Refinement types have also been developed for imperative languages. Flux~\cite{flux} adds liquid types to Rust's ownership type system,
allowing strong updates to change type predicates.
Refined TypeScript~\cite{refined_typescript} implements flow-sensitive refinement types for TypeScript by analyzing an SSA form%
---similar to \ranger's intermediate representation discussed in \autoref{sec:design}.
Scala has also been extended with qualified types~\cite{qualified_types_scala},
a form of refinement types that is compatible with the language's functional and object-oriented features.
Implementations of refinement types also exist for C~\cite{csolve} and Ruby~\cite{refinement_types_ruby}.

The Java ecosystem also features frameworks that support refinement types,
with an emphasis on language coverage and practical applicability.
The Checker Framework~\cite{java_checker_framework}
is a collection of extended typechecking plugins;
one of them is the Index Checker~\cite{java_index_checker}
which aims at verifying array indexing (and other collection accesses) using range-like constraints.
Recently, a form of liquid types has been implemented for Java~\cite{liquid_java},
with a focus on usability.
Our experimental evaluation in \autoref{sec:evaluation}
will compare \ranger to similar frameworks
in terms of expressiveness and degree of automation.

\subsubsection{Type inference.}

\ranger's type system implements a form of bidirectional typing~\cite{bidirectional_typing}, a technique that combines
top-down inference and bottom-up type-checking. As we detail in \autoref{sec:design}, \ranger's approach to bidirectional 
typing is however more global than the typical bidirectional typer, as we back-propagate types throughout every function.

\section{Motivating Example}
\label{sec:motivating-example}

\autoref{fig:motivating-example}
shows the most significant parts of a program that demonstrates
several of \ranger's capabilities.
It is written in \licorne (\ranger's current target language), %
whose syntax and semantics should be easy to glean
for readers familiar with languages like Kotlin and Scala
that combine functional and object-oriented features.

Function \Lic{decodeAll} in \autoref{ex:motivating:main}
implements the main functionality:
given an integer array \Lic{data},
\Lic{decodeAll} converts each of its elements
into a \Lic{Point}: a pair \Lic{x}, \Lic{y}
of integer coordinates
over an \Lic{xLen}-by-\Lic{yLen} bidimensional grid.
To this end,
\Lic{decodeAll} first filters the input \Lic{data}
to discard all negative integers;
then, it maps function \Lic{decode}
onto each element.
The actual conversion logic, implemented by \Lic{decode},
splits an integer into two parts:
the most significant bits are interpreted as the \Lic{x}
coordinate, whereas the 8 least significant bits become
the \Lic{y} coordinate.

\begin{figure}[!b]
  \centering
\begin{subfigure}{\textwidth}
\begin{lstlisting}[language=Licorne]
// Converts each integer in `data' to a bidimensional point over a `xLen'$\times$`yLen' grid.
fn decodeAll(data: Array[Int], xLen: <<1..>>, yLen: <<1..>>) -> List[Point[<<0..<xLen>>, <<0..<yLen>>]] {
  val validData = filter(data.toList(), fn (d) -> d >= 0);
  return map(validData, fn (d) -> decode(d, xLen, yLen));
}

// Converts `datapoint' to a bidimensional point over a `xLen'$\times$`yLen' grid.
fn decode(datapoint: <<0..>>, xLen: <<1..>>, yLen: <<1..>>) -> Point[<<0..<xLen>>, <<0..<yLen>>] {
  val x = clamp(0, xLen - 1, datapoint / 256);
  val y = datapoint % yLen;
  return new Point(x, y);
}

// Rounds `x' to the closest integer in the range <<min..max>>.
fn clamp(min: Int, max: <<min..>>, x: Int) -> <<min..max>> =
  when x <= min then min else when x <= max then x else max

record Point[I sub Int, J sub Int](x: I, y: J)
\end{lstlisting}
  \captionskip{Function \Lic{decodeAll} converts an array of integers encoding coordinates into instances of record type \Lic{Point}.}
  \label{ex:motivating:main}
\end{subfigure}

\begin{subfigure}{\textwidth}
\begin{lstlisting}[language=Licorne]
fn filter[T](l: List[T], p: pure fn(T) -> Bool) -> List[T with p(it)] {
  var rev = new Nil();
  for (var rem = l; rem is Cons; rem = rem.rest())
    { if p(rem.first()) { rev = new Cons(first=rem.first(), rest=rev); } };
  return reverse(rev);
}

fn map[T, U](ls: List[T], f: fn (T) -> U) -> List[U] { /* ... */ }
fn reverse[T](ls: List[T]) -> List[T] { /* ... */ }

class Array[T] {
  pure fn length(this) -> <<0..>> { /* ... */ }
  fn at(this, i: <<0..<this.length()>>) -> T  { /* ... */ }
  fn toList(this) -> List[T] {
    var ls = new Nil();
    for (var i = length() - 1; i >= 0; i -= 1) { ls = new Cons(first = at(i), rest = ls); };
    return ls;
  }
}
\end{lstlisting}
\captionskip{Auxiliary functions and classes used by \Lic{decodeAll} to process lists and arrays.}
  \label{ex:motivating:aux}
\end{subfigure}  

\caption{Motivating example.}
  \label{fig:motivating-example}
\end{figure}

\paragraph{Range types.}
\ranger's range refinement type annotations are \textcolor{rangercol}{highlighted in green}:
\begin{enumerate*}[label=\emph{\roman*})]
\item Type \Lic{<<1..>>}
  requires \Lic{xLen} and \Lic{yLen} to be positive integers,
  so that both dimensions are valid;
\item The \Lic{x} and \Lic{y}
  components of the \Lic{Point} instances returned by
  \Lic{decodeAll} (in a list)
  and \Lic{decode} must
  be respectively
  of type \Lic{<<0..<xLen>>} and \Lic{<<0..<yLen>>},
  that is integers between zero (included) and
  \Lic{xLen} or \Lic{yLen} (excluded);
\item Function \Lic{clamp} takes two integers
  \Lic{min} and \Lic{max} such that $\Lic{max} \geq \Lic{min}$
  (i.e., \Lic{max} has type \Lic{<<min..>>})
  and returns an integer in \Lic{<<min..max>>};
  \item The return type of \Lic{filter} applies the generic
    refinement operator \Lic{with} to parameter \Lic{T}
    to specify that the returned list only includes
    elements that satisfy filtering predicate \Lic{p}.
\end{enumerate*}

\autoref{fig:motivating-example}'s
type annotations define fundamental correctness conditions
of the program in a natural and succinct way.
Even though range types are not sufficient to express
full correctness, they still provide a rigorous documentation
and rule out several potential errors.
For example, typechecking the program verifies, among other things,
that the
\Lic{x} and \Lic{y} computed by \Lic{decode}
fall in the expected ranges;
and that \Lic{filter} returns a sublist
of nonnegative integers, which matches \Lic{decode}'s
requirement on its first argument \Lic{datapoint}.

\paragraph{Typechecking range types.}
Notably, \ranger can typecheck
the program against its typed signatures
without needing any annotations in the function \emph{bodies}.
To this end, \ranger deploys a bidirectional typechecking algorithm,
which \emph{infers} intermediate types in many cases.
For example, to typecheck \Lic{filter}'s body,
\ranger first back-propagates \Lic{filter}'s output type and records 
it as a type candidate for every assignment to \Lic{rev};
then, it performs a regular forward type analysis
which validates the candidates and successfully checks the program.
Typechecking \Lic{filter} also relies on 
\emph{smart casts}, a feature that makes \ranger flow-sensitive: 
since the update of \Lic{rev} inside the body
is guarded by condition \Lic{p(rem.first())},
\ranger can verify the conformance of the function to the predicate 
expressed by its return type: all elements of \Lic{rev}
satisfy predicate \Lic{p}.
This holds soundly only if \Lic{p} and \Lic{first} are 
pure functions. To this end,
\ranger includes a purity enforcement mechanism,
based on \Lic{pure} annotations such as the one in \Lic{p}'s typed signature: 
expressions used as type predicates
or as path conditions in smart casts
must be provably pure. 
Finally, typechecking
the loop in method \Lic{toList} of class \Lic{Array}
relies on another feature of \ranger: monotonicity analysis. 
The type-checker detects that \Lic{i} decreases 
across iterations, which, combined with smart cast
based on the loop's staying condition $\Lic{i}\geq 0$,
establishes that the array access is safe.

\paragraph{\ranger vs.\ other type systems.}
\ranger's refinement types can express properties that 
are out of reach for the type systems of mainstream 
languages with comparable features,
including Java, Kotlin, and Scala.
At the same time, as we detail in \autoref{sec:evaluation},
the amount of user-written annotations required by \ranger
is essentially the same as that of a regular Scala program.
In other words, programmers can ``upgrade'' to \ranger
without increasing their annotation burden.
\ranger is not the only current implementation
of refinement types for Java-like languages:
as discussed in \autoref{sec:related-work},
the Checker Framework and
Liquid Java provide comparable functionality for the Java language.
The experiments detailed in \autoref{sec:evaluation}
suggest that \ranger is often more flexible and expressive
than the Checker Framework; for example,
the latter cannot encode the constraint on \Lic{filter}'s return type
in \autoref{ex:motivating:aux}. 
While LiquidJava's annotation language is quite expressive,
it is limited in terms of language feature support;
in our experiments,
the LiquidJava implementation crashed on several examples
and failed to find bugs in programs that use
features like generics.

\section{Design and Implementation of a Range Refinement Type System}

The main \emph{design goals} of the \ranger range-focused refinement type
system are practicality and succinctness. 
This entails the following requirements:
\begin{enumerate*}[label=\emph{\roman*})]
\item \ranger's type system should be \emph{sound};
\item it should be sufficiently expressive to cover \emph{succinctly}
  the most \emph{common} programming idioms that involve ranges;
\item in these scenarios, it should be \emph{precise}
  without requiring user-written auxiliary annotations;
\item typechecking \ranger annotations should be efficient
  and deterministic.
\end{enumerate*}

\subsection{Features of \ranger}
\label{sec:features}

\paragraph{Refinements and ranges.}
\ranger's type system extends a standard type system
(similar to those available in languages like Scala and Kotlin)
with a \emph{refinement} type constructor $T\ \Lic{with}\ p(\Lic{it})$,
which denotes a subtype of $T$ consisting of all values $t \in T$
that also satisfy $p(t)$.
Predicate $p$ must be a pure boolean expression, that may depend on program values.
While the refinement constructor is generic, \ranger's focus is on integer ranges. 
To this end, we introduce a shorthand for integer range types: 
\Lic{<<l..u>>} denotes the range from \Lic{l} to \Lic{u} (both included), and 
\Lic{<<l..<u>>} the range from \Lic{l} (included) to \Lic{u} (excluded).
The notation also supports half-open intervals:
e.g., \Lic{<<0..>>} denotes the nonnegative integers.\footnote{
Our range notation is inspired by the one used for range expressions 
in Kotlin and Groovy.
}

\paragraph{Inference and analysis.}
\ranger should be usable with minimal auxiliary annotations.
Ideally, users would only annotate function signatures,
whereas the types of expressions within a function's body
should be inferred by the type checker.
To this end, \ranger implements a form 
of bidirectional typechecking: a backward type propagation 
runs \emph{before} typechecking to infer type 
\emph{candidates}, which are used in lieu of
explicit programmer-written annotations.
Type inference is best-effort and ``aggressive'': the 
typechecking phase is ultimately responsible for the 
soundness of typechecking, and will reject a type 
candidate if it cannot prove its validity or %
can infer a more precise type.

\paragraph{Precision and performance.}
To precisely check refinements, \ranger's typechecking
algorithm also relies on external constraint solving.
Since this dependency may significantly worsen performance,
\ranger pre-processes the constraints sent to the external solver
to make them tractable.
As we better discuss in \autoref{sec:implementation},
solver constraints are quantifier-free (linear) arithmetic; in
particular, 
analyzing loops does \emph{not} introduce fixpoint computations.

\paragraph{Practicality.}
Among other features that make it more flexible, 
\ranger includes a hybrid cast operator \Lic{!!}, useful
when statically checking a refinement is impractical.
For example, consider an array access \Lic{a.at(k)};
if \ranger cannot determine that \Lic{k} is in-bound
(e.g., because the access is guarded by a complex path condition),
we can replace \Lic{k} with \Lic{k!!}.
In this case, \ranger defers the check that \Lic{k} is in-bound
to run time, while still statically checking \Lic{k}'s base type.

\subsection{Notations and Preliminaries}
\label{sec:notation}

\begin{figure}[t]
  \centering
  \scriptsize
  \begin{alignat*}{3}
    \text{Program}~P  & ::= M^*
    \\
    \text{Method}~M  & ::= \IRc{fn}\ T_0.\IRc{fun}(v_1:T_1,\ldots,v_n:T_n) \to T_{\text{ret}}\ \{\ S\ \}
                       & %
    \\
    \text{Block}~S & ::= \epsilon\ \mid\ D\ \mid\ A\:S\ \mid\ N\:S
    \\
    \text{Control-flow}~D & ::= \IRc{if}\ c\ \IRc{then}\ S\ \IRc{else}\ S\ \IRc{merge}\ J^*\ \IRc{next}\ \ S
                          & c \in \textsl{Vars} \\
                      & \quad\mid \IRc{from}\ J^*\ \IRc{while}\ S (c)\ \IRc{loop} \ S\ \IRc{next}\ S\ \mid\ \IRc{return}\ v \mid \IRc{panic}\ v
                        & c, v \in \textsl{Vars}
    \\
    \text{Phi}~J          & ::= v_{\text{out}} := \phi(v_{\text{left}}, v_{\text{right}})
                          & v_{\text{out}}, v_{\text{left}}, v_{\text{right}} \in \textsl{Vars}
    \\
    \text{Assignment}~A & ::= v\ \IRc{:=}\ e\
                        & v \in \textsl{Vars}
    \\
    \text{Type ascription}~N & ::= v :: T
                         & v \in \textsl{Vars}
    \\
    \text{Type}~T  & ::= B \mid B\ \IRc{with}\ q \mid T \intertype T \mid T \uniontype T
    \\
    \text{Base type}~B & ::= \IRc{Any} \mid \IRc{Nothing} \mid \IRc{Unit} \mid \IRc{Int} \mid \IRc{Bool} \mid \IRc{String} \mid C
                       & C \in \textsl{Classes}
    \\
    \text{Predicate}~q & ::= p \mid l
                       & p \in \textsl{Preds}
    \\
    \text{Expression}~e & ::= v \mid v\ \IRc{as}\ B\ \mid\ v\IRc{!!} \mid c \mid l \mid a \mid \IRc{unit}
                    & v \in \textsl{Vars}
    \\
    \text{Call expression}~c & ::= \IRc{call}\ v_0.f(v_1,\ldots,v_n)
                    & v_k \in \textsl{Vars}
    \\
    \text{Logic expression}~l & ::= \IRc{true} \mid \IRc{false} \mid v_1 \diamond v_2 \mid v_1 \bowtie v_2 \mid {!}v_0 \mid v_0\ \IRc{is}\ B
                    & v_k \in \textsl{Vars}\quad {\diamond} \in \{{\land}, {\lor}\}\quad {\bowtie} \in \{{<}, {\leq}, {=}\}
    \\
    \text{Arithmetic expr.}~a & ::= n \mid v_1 \boxdot v_2
                    & n \in \mathbb{Z}\quad v_k \in \textsl{Vars}\quad \boxdot \in \{{+}, {-}, \times, {/}, \% \}
  \end{alignat*}
  \caption{The syntax of \ircorne, \licorne's intermediate representation.
    \textsl{Vars} stands for the set of program variables,
    each corresponding to a uniquely defined value in SSA form;
    \textsl{Preds} for the set of predicate identifiers (pure functions that return \Lic{Bool}),
    \textsl{Classes} for the set of user-defined types,
    and $\mathbb{Z}$ for the set of integer numbers. \Lic{Nothing} and \Lic{Any} are the bottom and top types.}
  \label{fig:syntax}
\end{figure}

\begin{figure}[!tbh]
  \centering
  \begin{subfigure}{0.35\linewidth}
\begin{lstlisting}[language=Licorne,numbers=none]
fn maxPos(a: Array[Int]) -> <<0..>> {
  var k = 0; var max = 0;
  while (k < a.length()) {
    var v = a.at(k);
    if (v > 0 && v > max) {
      max = v;
    };
    k += 1;
  };
  return max;
}
\end{lstlisting}
    \caption{Function \Lic{maxPos} in \licorne.}
    \label{ex:ir:source}
  \end{subfigure}
  \hspace{2mm}
  \begin{subfigure}{0.6\linewidth}
\begin{lstlisting}[language=IRcorne,numbers=none]
fn maxPos(a: Array[Int]) -> <<0..>> {
  k$_0$ := 0 ; max$_0$ := 0
  from k$_1$:=$\phi$(k$_0$,k$_2$) ; max$_1$:=$\phi$(max$_0$,max$_3$)
  while tmp$_0$ := call a.length() ; c$_0$ := k$_1$ < tmp$_0$ (c$_0$) loop
    v$_0$ := call a.at(k$_1$)
    tmp$_1$ := v$_0$ > 0 ; tmp$_2$ := v$_0$ > max$_1$ ; c$_1$ := tmp$_1$ $\land$ tmp$_2$
    if c$_1$ then max$_2$ := v$_0$ else $\epsilon$
    merge max$_3$ := $\phi$(max$_2$,max$_1$)
    next k$_2$ := k$_1$ + 1
  next return max$_1$
}
\end{lstlisting}
    \caption{Function \Lic{maxPos} in the \ircorne SSA intermediate representation.}
    \label{ex:ir:ircorne}
  \end{subfigure}
  \caption{Function \Lic{maxPos} returns the largest non-negative element in array \Lic{a}, or zero if all elements of \Lic{a} are negative.}
  \label{fig:ir-example}
\end{figure}

\subsubsection{IR syntax.}
We present \ranger's type system on programs 
written in \ircorne: the intermediate representation
that follows \autoref{fig:syntax}'s syntax.
This both makes the presentation more accessible
and matches the actual implementation of \ranger,
which also works on a (more complex) IR.
\ircorne is an SSA form,
where every variable is assigned at most once
and $\phi$ functions merge values from different control-flow paths.
Unlike classic SSA representations, however,
\ircorne includes explicit structural control flow instructions
(i.e., loops and conditionals),
which are leveraged by \ranger's flow-sensitive analyses.
\autoref{fig:ir-example} shows the \ircorne translation
of a simple function \Lic{maxPos} that computes the largest 
positive element in its input array \Lic{a},
which demonstrates \ircorne's syntax in practice.

\ircorne
programs are collections of methods, whose body is a (possibly empty $\epsilon$)
sequence of instructions
(i.e., a block, which also defines a scope).
Instructions include
assignments and
ascriptions (checks that a variable has a given type),
as well as control-flow instructions:
conditionals, loops, returns, and \IRc{panic}
(terminate abruptly with an error).
A conditional \IRc{if$\,c\,$then$\,t\,$else$\,e\,$merge$\,m\,$next$\,n$}
introduces two blocks $t$ and $e$, one of which executes
depending on the Boolean condition $c$;
in addition,
a sequence $m$ of $\phi$ functions merge the variables defined in the two branches;
and $n$ declares the block executed afterwards.
A loop \IRc{from$\,m\,$while$\,w(c)\,$loop$\,b\,$next$\,n$}
executes a block $b$ as long as a condition $c$ is true;
correspondingly, $m$ is a sequence of $\phi$ functions that merge
the variables in any predecessor block
(including the back-edge from the loop's previous iteration);
$w$ are instructions that evaluate the condition $c$;
and $n$ determines the block executed after the loop exits.
The \IRc{next} blocks make \ircorne a \emph{functional} SSA form~\cite{DBLP:journals/sigplan/Appel88},
where the targets of branching instructions are explicitly available.

\subsubsection{Notations.}
We use the symbols $\uniontype$ and $\intertype$ for union and intersection types;
$T_1 \subtype T_2$ denotes that $T_1$ is a subtype of $T_2$.
We also use $\rep{A_k}{k}$ to denote the sequence $A_1\,A_2\,\ldots$ of $A_k$ repeated for every value of $k$.

\subsection{How \ranger Works}
\label{sec:design}

\subsubsection{Flattening.}
\label{subsec:ff-proc}
A disadvantage of the SSA form %
is that it breaks down complex expressions into assignment sequences,
which complicates analyses that need whole source expressions.
To address this issue,
we record the origin of each expression and make this information
available to \ranger's analysis,
which can use it to recover the \emph{flat} form of every variable $v$,
i.e., the complete expression whose value is stored by $v$.
For 
instance, the flat form of $\IRc{c}_1$ in \autoref{ex:ir:ircorne}
is $\IRc{v}_0 > \IRc{0}\ \land\ \IRc{v}_0 > \IRc{max}_1$.

\subsubsection{Candidate types inference.}
\ranger performs a backwards analysis that propagates types known from ascriptions, declared method return types, and call arguments, 
to help the subsequent typechecking phase with type synthesis. 
Formally, candidate types inference is a modular backward dataflow analysis, run individually on every method of the program.
For each instruction $i$,
the analysis computes a mapping $\cplain{i} \colon \textsl{Vars} \to 2^T$ such that $\cplain{i}(v)$
is the set of inferred types of variable $v$ before $i$.
Concretely, the dataflow analysis tracks
the inferred types after ($\cout{i}$)
and before ($\cin{i} = \cplain{i}$) every $i$.
The rules in \autoref{tab:candidate-backward-rules}
define how $\cin{i}$ is updated working backwards from $\cout{i}$:
\begin{enumerate*}[label=\emph{\roman*})]
\item If $i$ is an exit point (rule \rulen{exit}),
  $\cout{i}(x)$ is the empty set,
  which carries no information about $x$'s type;
  then, $\cin{i}$ uses the declared return type $T_{\text{ret}}$
  as the candidate type of the returned variable.
\item If $i$ is a call (rule \rulen{call}),
  the analysis tries to resolve the called method $m$,
  and then uses its declared parameter types $T_k$
  as new candidates for the types of the actual argument variables;
  if $m$ cannot be resolved, no candidates are added.
\item If $i$ is an assignment of $y$ to $x$ (rule \rulen{assign}),
  $\cin{i}(y)$ is updated based on the inferred type $\cout{i}(x)$
  of $x$.
\item If $i$ joins two variables with a $\phi$ function
  (rule \rulen{join}),
  the inferred type $\cout{i}(v)$ of the unified variable
  is propagated to both source variables.
\item In all other cases,
  $\cout{i}(x)$ is the union of all candidate types inferred in the
  \emph{successor} nodes $j$ of $i$;
  and $\cin{i} = \cout{i}$, as the inferred candidates are simply backward propagated.
\end{enumerate*}

\begin{table}[!tb]
  \centering
  \scriptsize
    \setlength{\tabcolsep}{5pt}
    \begin{tabular}{lcll}
      \multicolumn{1}{c}{\textsc{rule}} & \multicolumn{1}{c}{\textsc{instruction} $i$} & \multicolumn{1}{c}{$\cin{i}$} & \multicolumn{1}{c}{$\cout{i}(x)$} 
      \\
      \midrule
      \rulen{exit} & $\IRc{return}\ v$ & $\cout{i}[v \mapsto \{T_{\text{ret}}\}]$ & $\emptyset$
      \\
      \rulen{call} & $r\ \IRc{:=}\ \IRc{call}\ v_0.m(v_1, \ldots)$
      & $\cout{i}\rep{[v_k \mapsto \cout{i}(v_k) \cup \{ T_k \}]}{k}$
      & \multirow{4}{*}{$\bigcup_{j \in \textsf{succ}(i)} \cin{j}(x)$}
      \\
      \rulen{assign} & $x$ := $y$
      & $\cout{i}[y \mapsto \cout{i}(y) \cup \cout{i}(x)]$
      \\
      \rulen{join} & $v := \phi(x_1, x_2)$
      & $\cout{i}\rep{[x_k \mapsto \cout{i}(x_k) \cup \cout{i}(v)]}{k=1,2}$
      \\
        \rulen{skip} & any other instruction
      & $\cout{i}$
    \end{tabular}
    \caption{Candidate types inference rules.
      Each rule defines the set $\cin{i}$ of candidate inferred types
      before instruction $i$ in terms of the set $\cin{i}$
      that hold after instruction $i$.}
  \label{tab:candidate-backward-rules}
\end{table}

In \autoref{fig:ir-example}'s example,
\ranger's type inference algorithm assigns type \IRc{<<0..>>}
to every variant \IRc{max$_k$} of variable \IRc{max}.
Consider the control flow of the instructions involving \IRc{max}:
\noindent\makebox[\linewidth][c]{
  \begin{tikzpicture}[start chain=going right, node distance=4mm]
    \begin{scope}[every node/.style={on chain}]
    \node (ret) {$\Lic{return max}_1$};
    \node  (m0) {$\Lic{max}_0 :=$};
    \node  (m1) {$\Lic{max}_1 :=$};
    \node  (m2) {$\Lic{max}_2 :=$};
    \node  (m3) {$\Lic{max}_3 :=$};
  \end{scope}
  \begin{scope}[->,>=stealth,thick]
    \draw (m0) -- (m1);
    \draw (m1) -- (m2);
    \draw (m2) -- (m3);
    \draw (m1) to[bend left=10] (m3);
    \draw (m3) to[bend left=10] (m1);
    \draw (m1) to[bend right=10] (ret);
  \end{scope}
\end{tikzpicture}
}
The analysis starts at the \IRc{return} exit point,
where it infers type \IRc{<<0..>>} for $\IRc{max}_1$;
from there, it goes backwards
into the loop's head through the $\phi$ assignment to $\IRc{max}_1$,
which infers the same type for $\IRc{max}_0$ and $\IRc{max}_3$;
and then traverses the loop body
until it reaches the $\phi$ assignment to $\IRc{max}_3$,
where it infers \IRc{<<0..>>} also for $\IRc{max}_2$.

\subsubsection{Monotonicity detection.}
\label{subsubsec:monotonicity-detection}
\ranger analyzes loops to detect integer variables that are updated \emph{monotonically} across loop iterations.
The output of this analysis 
is a partial mapping $M \colon \textsl{Vars} \to \textsl{Ranges}$
of variables to range types.
As discussed later, \ranger's type analysis uses this information
to refine the inferred type of loop variables.
Formally, monotonicity detection processes every $\phi$ assignment
$v_{\text{next}}\ \IRc{:=}\ \phi(v_{\text{init}}, v_{\text{prev}})$
in a loop's \IRc{from} clause,
which merges the values of variable $v$ set before the loop ($v_{\text{init}}$)
and during the previous iteration ($v_{\text{prev}}$).
\ranger constructs two monotonicity constraints:
$\mu_{\uparrow} \triangleq f(v_{\text{prev}}) \geq v_{\text{next}}$
and 
$\mu_{\downarrow} \triangleq f(v_{\text{prev}}) \leq v_{\text{next}}$,
where $f(v_{\text{prev}})$ denotes the \emph{flat} form of $v_{\text{prev}}$
described above.
\ranger uses an SMT solver to check the validity of the monotonicity
constraints:
if $\mu_{\uparrow}$ is valid, $v_{\text{next}}$ is monotonically nondecreasing,
and hence $M(v_{\text{next}})$ is the type $\Lic{<<}v_{\text{init}}\Lic{..>>}$;
if $\mu_{\downarrow}$ is valid, $v_{\text{next}}$ is monotonically nonincreasing,
and hence $M(v_{\text{next}})$ is the type $\Lic{<<..}v_{\text{init}}\Lic{>>}$;
if both are valid, $v_{\text{next}}$ is constant,
and hence $M(v_{\text{next}})$ is the type $\Lic{<<}v_{\text{init}}\Lic{..}v_{\text{init}}\Lic{>>}$.
In all other cases (including if the flat form is unavailable),
$M(v_{\text{next}})$ is undefined,
so that monotonicity detection will always be sound.

In \autoref{fig:ir-example}'s example, 
\ranger determines that $\IRc{k}_1$---set in the loop's head---%
is monotonically non-de\-creas\-ing:
the flat form of $\IRc{k}_2$ is $\IRc{k}_1 + 1$,
and
the constraint $\mu_{\uparrow} \triangleq \IRc{k}_1 + 1 \geq \IRc{k}_1$
is obviously valid;
hence, $\IRc{k}_1$ has type \Lic{<<k$_0$..>>}.

\begin{figure}
    \centering
    \scriptsize
\begin{subfigure}{1.0\linewidth}
    \begin{mathpar} %

    \inferrule[function]{
      \Gamma, \rep{v_k \colon T_k}{k} \vdash S \returns T_{\text{ret}}, \top
    }{
      \Gamma \vdash \IRfn\ T_0.\IR{fun}(v_1\colon T_1,\ldots) \to T_{\text{ret}}\ \{\ S\ \}
    }

    \inferrule[assignment]{
      \Gamma \vdash e \colon  T ; \psi^+|\psi^- \\
      \Gamma, v\colon T ; \psi^+|\psi^- \vdash S \returns T_{\text{ret}}, \tau
    }{
      \Gamma \vdash v\ \IR{:=}\ e\ ;\ S \returns T_{\text{ret}}, \tau
    }

    \inferrule[empty]{}{
      \vdash \epsilon \returns \IRNothing, \bot
    }\hspace{-4mm}

    \inferrule[ascription]{
     \Gamma \vdash v \colon V \\
     \Gamma, v \colon V \vdash S \returns T_{\text{ret}}
    }{
     \Gamma \vdash v :: V\ ;\ S \returns T_{\text{ret}}
    }\hspace{-4mm}
    
    \inferrule[return]{
      \Gamma \vdash v \colon  T
    }{
      \Gamma \vdash \IRreturn\ v \returns T, \top
    }\hspace{-4mm}

    \inferrule[panic]{
      \Gamma \vdash\ v \colon  \IR{String}
    }{
      \Gamma\ \vdash\ \IRpanic\ v\ \returns \IRNothing, \top
    }

    \inferrule[loop]{
      \rep{\Gamma \vdash v_{k, \text{init}} \colon  T_k}{k} \\
      \rep{\Gamma \vdash v_{k, \text{prev}} \colon  T_k}{k} \\
      \Gamma, \rep{v_{k, \text{next}} \colon  T_k}{k} \vdash W \returns T_{\text{ret}}, \tau_W \\
      \Gamma, \rep{v_{k, \text{next}} \colon  T_k}{k} \vdash c \colon  \IRBool ; \psi^+|\psi^- \\
      \Gamma, \psi^+, \rep{v_{k, \text{next}} \colon  T_k}{k} \vdash S \returns T_{\text{ret}}, \tau_{S} \\
      \Gamma, \psi^-, \rep{v_{k, \text{next}} \colon  T_k}{k} \vdash N \returns T_{\text{ret}}, \tau_{N}
    }{
      \IRfrom\
      \rep{v_{k, \text{next}} \IR{:=} \phi(v_{k,  \text{init}}, v_{k, \text{prev}})}{k}
      \IRwhile\ W (c)\
      \IRloop\ S\
      \IRnext\ N\ 
      \returns T_{\text{ret}}, \tau_{W} \lor \tau_{N}
    }

    \inferrule[conditional]{
      \Gamma \vdash c \colon  \IRBool ; \psi^+|\psi^- \\
      \Gamma, \psi^+ \vdash S_{\text{then}} \returns T_{\text{ret}}, \tau_{\text{then}} \\
      \Gamma, \psi^- \vdash S_{\text{else}} \returns T_{\text{ret}}, \tau_{\text{else}} \\
      \rep{\Gamma \vdash v_{k, \text{then}} \colon  T_k}{k} \\
      \rep{\Gamma \vdash v_{k, \text{else}} \colon  T_k}{k} \\
      \Gamma, \rep{v_{k} \colon T_k}{k}, \inclif{\psi^+}{\tau_{\text{else}}}, \inclif{\psi^-}{\tau_{\text{then}}} \vdash N \returns T_{\text{ret}}, \tau_{\text{after}} \\
      \tau = (\tau_{\text{then}} \land \tau_{\text{else}}) \lor \tau_{\text{after}}
    }{
      \Gamma\ \vdash\ \IRif\ c\ \IRthen\ S_{\text{then}}\ \IRelse\ S_{\text{else}}\ \IRmerge\ \rep{v_{k} \IR{:=} \phi(v_{k, \text{then}},v_{k, \text{else}})}{k}\ \IRnext\ N \returns T_{\text{ret}}, \tau
    }

  \end{mathpar}
  \caption{Instructions and control flow.}
  \label{fig:typing-rules-control}
\end{subfigure}

\begin{subfigure}{1.0\linewidth}
    \begin{mathpar} %

    \inferrule[subtype]{
      \Gamma \vdash v \colon T_1 \\
      \Gamma \vdash T_1 \subtype T_2
    }{
      \Gamma \vdash v \colon T_2
    }

    \inferrule[call]{
      \Gamma \vdash \rep{v_k \colon T_k}{k} \\
      \IR{f} \colon \rep{U_k}{k} \to T_{\text{ret}} \\
      \rep{T_k \subtype U_k}{k}
    }{
      \Gamma \vdash \IRcall\ v_0.\IR{f}(v_1,\ldots)\ \colon\ T_{\text{ret}}
    }

    \inferrule[var]{
      v \colon T \in \Gamma
    }{
      \Gamma \vdash v \colon T
    }

    \inferrule[hybrid-cast]{
      \Gamma \vdash v \colon B
    }{
      \Gamma \vdash v\IRbb\: \colon B\ \IRwith\ p \wr p(v)
    }

    \inferrule[cast]{
      \Gamma \vdash v \colon B_1\\
      \Gamma \vdash B_2 \subtype B_1
    }{
      \Gamma \vdash v\ \IRas\ B_2 \colon B_2
    }

    \inferrule[and]{
      \Gamma \vdash b_1 \colon \IRBool ; \psi_{1}^+|\psi_{1}^- \\
      \Gamma \vdash b_2 \colon \IRBool ; \psi_{2}^+|\psi_{2}^-
    }{
      \Gamma \vdash b_1 \land b_2 \colon \IRBool\ ;\ \psi_{1}^+, \psi_{2}^+\ |\ \emptyset
    }

    \inferrule[or]{
      \Gamma \vdash b_1 \colon \IRBool ; \psi_{1}^+|\psi_{1}^- \\
      \Gamma \vdash b_2 \colon \IRBool ; \psi_{2}^+|\psi_{2}^-
    }{
      \Gamma \vdash b_1 \lor b_2 \colon \IRBool\ ;\ \emptyset\ |\ \psi_{1}^-, \psi_{2}^-
    }

    \inferrule[not]{
      \Gamma \vdash b \colon \IRBool ; \psi^+|\psi^-
    }{
      \Gamma \vdash \IR{!}b \colon \IRBool ; \psi^-|\psi^+
    }

    \inferrule[type-test]{
      \Gamma \vdash b \colon B
    }{
      \Gamma \vdash b\ \IRis\ B \colon \IRBool\ ;\ b \colon B\ |\ \emptyset
    }

    \inferrule[comparison]{
      \Gamma \vdash a_1 \colon \IR{Int} \\
      \Gamma \vdash a_2 \colon \IR{Int}
    }{
      \Gamma \vdash a_1 \bowtie a_2 \colon \IRBool\ ;\ a_1 \bowtie a_2\ |\ \lnot \bigl(a_1 \bowtie a_2\bigr)
    }

    \inferrule[int-literal]{
      n \in \mathbb{Z}
    }{
      n \colon \{n\}
    }

    \inferrule[bool-literal]{
     b \in \{\IR{true}, \IR{false}\}
    }{
     \vdash b \colon \IRBool
    }

    \inferrule[unit-literal]{}{
     \vdash \IR{unit} \colon \IRUnit
    }

    \inferrule[arith]{
      \Gamma \vdash a_1 \colon I_1 \subtype \IRInt \\
      \Gamma \vdash a_2 \colon I_2 \subtype \IRInt
    }{
      \Gamma \vdash a_1 \boxdot a_2 \colon I_1 \odot I_2
    }

  \end{mathpar}
  \caption{Expressions.}
  \label{fig:typing-rules-expressions}
\end{subfigure}

\begin{subfigure}{1.0\linewidth}
    \begin{mathpar}

    \inferrule[refinement]{
      \Gamma \vdash v \colon B \\
      \Gamma \vdash p(v)
    }{
      \Gamma \vdash v \colon B\ \IRwith\ p(\IRit)
    }
      
    \inferrule[sub-ref]{
      \Gamma \vdash B_1 \subtype B_2 \\
      \Gamma \vdash q_1 \Rightarrow q_2
    }{
      \Gamma \vdash B_1\ \IR{with}\ q_1 \subtype B_2\ \IR{with}\ q_2
    }\hspace{-2mm}

    \inferrule[mono]{
      \Gamma \vdash v : I \subtype \IR{Int} \\
      M(v) = \IRlc a \IRdd b \IRrc
    }{
      \Gamma \vdash v : I \intertype M(v)
    }
    \end{mathpar}
    \caption{Subtyping relation for refinement types,
      and usage rules for the monotonicity analysis.}
    \label{fig:subtyping}
    \label{fig:monotonicity-app-rules}
\end{subfigure}
\caption{\ranger's typing rules.}
\label{fig:typing-rules}
\end{figure}

\subsubsection{Type checking.}
\autoref{fig:typing-rules} shows \ranger's typing rules,
which we illustrate in the following.
The rules consist of typing judgments $\vdash$ using an environment
$\Gamma$ that is a sequence of pairs $v \colon T$%
---denoting that variable $v$ has type $T$.

\paragraph{Instructions.}
The type rules for instructions in \autoref{fig:typing-rules-control}
use typing judgments
$\Gamma \vdash S \returns T, \tau$,
denoting that
all \IRc{return} instructions in block $S$ return a value of type $T$;
$\tau$ is a Boolean that is $\top$ iff all paths through $S$
end with a terminating instruction.
For readability,
we omit $\tau$ in rules that are not affected by it.
Thus,
rule \rulen{function} simply reduces checking a function to checking its body.
Rules \rulen{return}, \rulen{panic}, and \rulen{empty}
establish the typing judgments at the terminal points of each path.
Rule \rulen{ascription} expresses an ascription as a type check.
Rule \rulen{assignment} binds the type of source expression $e$
\emph{and} the associated control-flow information to the assignment target $v$,
while forwarding return type $T_{\text{ret}}$ and termination information $\tau$.

\paragraph{Branching.}
The rules for \rulen{loop}s and \rulen{conditional}s
adapt \cite{DBLP:conf/icfp/Tobin-HochstadtF10}'s
formalization of occurrence typing to model smart casts:
$c : \IRc{Bool};\psi^+|\psi^-$ means
that $\psi^+$ holds if $c$ is true, and $\psi^-$ holds if $c$ is false. 
Rule \rulen{loop}
uses this information to check loops:
\begin{enumerate*}[label=\emph{\roman*})]
\item the type $T_k$ of variables $v_{k,\text{init}}$ and $v_{k,\text{prev}}$
  that are merged at the loop's head should be the same;
\item correspondingly, $\phi$-merged variable $v_{k,\text{next}}$
  also has the same type $T_k$, a piece of information that is used to check
  the \IRc{while} block $W$ and to type the condition itself $c$;
\item the loop body $S$ is checked under $\psi^+$,
  which holds when $c$ is true (staying condition);
\item conversely, the next block $N$ is checked under the complementary
  $\psi^-$ (exit condition);
\item any return points in $W$, $S$, and $N$ should return the same
  type $T_{\text{ret}}$ of the whole loop;
\item finally, the whole instruction block always terminates the function
  if \IRc{while} block $W$ or the \IRc{next} block
  always does (i.e., $\tau_W \lor \tau_N$).
\end{enumerate*}

Rule \rulen{conditional} works in a similar fashion:
\begin{enumerate*}[label=\emph{\roman*})]
\item $\psi^+$, which holds when condition $c$ is true,
  is used to check the $S_{\text{then}}$ block,
  whereas the complement condition $\psi^-$ is used to check the $S_{\text{else}}$ block;
\item the type $T_k$ of variables $v_{k,\text{then}}$ and $v_{k,\text{else}}$
  that are merged after the conditional should be the same,
  and is also propagated to the $\phi$-merged variable $v_{k}$;
\item typing the next block $N$ can use condition $\psi^+$ iff $\tau_{\text{else}}$,
  and condition $\psi^-$ iff $\tau_{\text{then}}$:
  in fact, if the \IRc{else} block terminates the enclosing function ($\tau_{\text{else}} = \top$),
  only the \IRc{then} condition $\psi^+$ holds after the conditional,
  and vice versa for $\psi^-$ and $\tau_{\text{then}}$;
\item finally, the whole instruction block always terminates the function
  (i.e., $\tau = \top$)
  if both the \IRc{then} and \IRc{else} always terminate
  (i.e., $\tau_{\text{then}} \land \tau_{\text{else}}$)
  or the \IRc{next} block always does (i.e., $\tau_{\text{after}} = \top$).
\end{enumerate*}

\paragraph{Expressions.}
\autoref{fig:typing-rules-expressions} presents the main typing rules for expressions, most of which are straightforward:
\begin{enumerate*}[label=\emph{\roman*})]
\item \rulen{subtype} and \rulen{var} formalize the usual subtyping and typing
  relations;
\item \rulen{call} models a qualified call,
  where the callee \Lic{f}'s signature is resolved based on the target $v_0$'s type;
\item \rulen{int-literal}, \rulen{bool-literal}, and \rulen{unit-literal}
  are the obvious typing rules for literal of different types.
\end{enumerate*}
The other expression rules are specialized to match \ranger's
flow-sensitivity:
\begin{enumerate*}[label=\emph{\roman*})]
\item \rulen{and}, \rulen{or}, and \rulen{not} augment
  the usual rules for Boolean expressions
  with the propagation of the flow conditions $\psi^+, \psi^-$;
  for example, \rulen{and} shows that the condition that holds when $b_1 \land b_2$
  is true is the union of $\psi_1^+$ and $\psi_2^+$,
  since both $b_1$ and $b_2$ are true;
\item similarly, \rulen{type-test} and \rulen{comparison}
  augment the usual rules for type tests and comparison expressions
  with an encoding of their flow conditions;
\item \rulen{arith} captures a simple form of abstract interpretation
  over the interval domain~\cite{CousotC77}, which derives
  a sound approximation of the range of an expression from its operands' ranges:
  $\odot$ is an abstract version of the arithmetic operations
  over this abstract domain;
  for example, $\Lic{<<}a\Lic{..}b\Lic{>>} \oplus \Lic{<<}c\Lic{..}d\Lic{>>}$
  is range $\Lic{<<}a+c\Lic{..}b+d\Lic{>>}$.
\end{enumerate*}

\paragraph{Casts and hybrid casts.}
\autoref{fig:typing-rules-expressions}
also displays the typing rules for casts.
Rule \rulen{cast} models a normal cast
that narrows the type of $v$ to a \textit{base} subtype of its currently known type.
In contrast, rule \rulen{hybrid} formalizes the hybrid cast operator $\IRc{!!}$,
which provides an unsound escape hatch for the static typing rules.
The rule allows the type checker to
\emph{assume} that $v\IRc{!!}$ has refinement type $B\ \IRc{with}\ p(\IRc{it})$
but only checks statically that $v$ has base type $B$;
the check that $p(v)$ holds is deferred to run time
(denoted $\wr\: p(v)$ in the rule).
The term ``hybrid cast'' is taken from~\cite{DBLP:conf/popl/Flanagan06},
where it denotes a similar feature.

\paragraph{Refinements rules.}
\autoref{fig:subtyping} presents the subtyping rules related to refinement types.
Rule \rulen{refinement} encodes the semantics of the \IRc{with} refinement
construct.
Then, 
rule \rulen{sub-ref}
reduces the subtyping relation between two refinement types
to subtyping between their base types
\emph{and}
implication between their refinement predicates.
In particular,
rule \rulen{refinement} implies
that any type $T$ is equivalent to $T\ \IRc{with}\ \IRc{true}$;
and rule \rulen{sub-ref}%
---evaluated with $q = q_1$, $B = B_1 = B_2$, and $q_2 = \IRc{true}$---%
implies that $B\ \IRc{with}\ q \subtype B$ for any predicate $q$.
We omit \ranger's other subtyping rules (esp.\ those for union and intersection types), as they are very similar to those of well-known type systems such as Scala's.

\paragraph{Monotonicity.}
Rule \rulen{mono} in \autoref{fig:monotonicity-app-rules}
illustrates how \ranger's type checking
uses the monotonicity information collected in a previous analysis.
For every integer loop variable $v$
that has been shown to be monotonically nondecreasing
(range \Lic{<<a..>>}),
nonincreasing
(range \Lic{<<..b>>},
or constant
(range \Lic{<<a..a>>}),
\ranger narrows the type of $v$ by intersecting
it with the corresponding range type.

\subsubsection{Using type candidates to help typechecking.}
Following the \emph{bidirectional} typechecking paradigm~\cite{bidirectional_typing},
\ranger uses the type candidates $\cand$
as \emph{heuristics} to help type analysis in two phases.
First, it uses type candidates to sift through the predicates introduced by smart casting:
in fact,
smart casting predicates reflect a program's path conditions,
and hence they can become quite complex;
as a result,
using all predicates to refine integer-typed expressions may
proliferate the number of implications to check,
which can slow down type checking considerably.
Instead, \ranger's type checking algorithm
uses a few heuristics to select which smart casting predicates
to use;
in particular,
it prefers
predicates that determine subtypes of any of the candidate types.
Second,
when \ranger's type checking algorithm processes a loop,
it needs to guess the types $T_k$  of the variables $v_k$
merged in the \IRc{from} $\phi$-function block;
in particular, the type of the $v_{k, \text{prev}}$ variables
is unknown initially, since type checking is a forward analysis.
To this end,
\ranger selects one of the available type candidates $C \in \cand(v_{k, \text{prev}})$
at the loop's entry,
checks that $C$ is compatible
with the type of the other merged variable $v_{k, \text{init}}$,
and provisionally assumes that $v_{k, \text{prev}} \colon C$;
when reaching the loop body's end,
it will have enough information about $v_{k, \text{prev}}$
to finally validate the assumption.
If validation fails, typechecking will also fail:
the inferred information was insufficient,
and hence additional annotations are required.
In all,
the type checking algorithm
ensures soundness,
so that type inference only provides best effort suggestions.

\paragraph{Example.}
In \autoref{fig:ir-example}'s example, \ranger first applies
rules \rulen{int-literal} and \rulen{assignment} to determine $\IRc{k}_0, \IRc{max}_0 \colon \{ \IRc{0} \}$.
Rule \rulen{loop} uses the loop condition to smartcast $\IRc{k}_1$ to $\IRc{<<..<a.length()>>}$, while 
monotonicity analysis strengthens $\IRc{k}_1$'s type to $\IRc{<<0..<a.length()>>}$,
ruling out out-of-range errors in the array access $\IRc{a.at(k}_1\IRc{)}$.
Furthermore, when starting to analyze the loop, \ranger guesses the type $\IRc{<<0 ..>>}$ for $\IRc{max}_1$ 
(return type back-propagated as a type candidate). 
Thanks to rule \rulen{conditional} and the condition that guards the assignment of $\IRc{max}_2$, 
\ranger derives $\IRc{max}_2 \colon \IRc{<<1 ..>>}$ and $\IRc{max}_3 \colon \IRc{<<0 ..>>}$, the latter 
validating the guess.

\paragraph{Purity.}
Predicates appearing in refinement types must be \emph{pure} functions.
\ranger includes a sound purity analysis for methods and closures,
which also works at the level of \ircorne.
In short, a \Lic{pure} function can only call other pure functions,
and compute and return the value of pure expressions---possibly
split into intermediate assignments.

\paragraph{Other features.}
For space constraints, we cannot discuss in detail how
\ranger is not limited to basic imperative constructs but integrates
well with \licorne's object-oriented and functional features,
including classes, enum-like datatypes, and constrained genericity.
For example, the return type of \autoref{fig:motivating-example}'s function \Lic{filter}
depends on argument \Lic{p}, used as a predicate to refine a type parameter.
\ranger also (soundly) relaxes \licorne's method overriding rules:
while a method's argument types must be invariant
in \licorne (like in Java or Kotlin), %
\ranger allows
an overriding redefinition of an argument of type $\tau\ \Lic{with}\ p$
to be invariant in $\tau$ but contravariant in $p$.

\subsection{Implementation}
\label{sec:implementation}

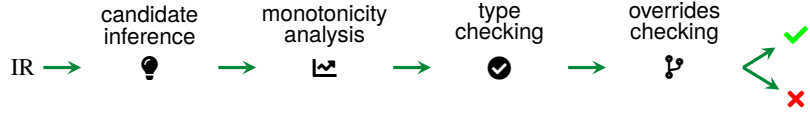
\begin{figure}[!bt]
  \centering
\begin{tikzpicture}[
  step node/.style={
    rectangle, %
    inner sep=1pt, fill=white,
    minimum width=18mm,
    label={[align=center,font=\sffamily]above:#1}
  },
  connect/.style={
    ->,
    >=stealth,
    very thick,
    rangercol
  },
  node distance=7mm and 5mm
  ]
  
  \node[align=center,font=\scshape] (IR) {IR};

  \node[right=of IR,
  step node={candidate\\[-3pt]inference}] (infer) {\faLightbulb};
  \node[right=of infer,
  step node={monotonicity\\[-3pt]analysis}] (mono) {\faChartLine};
  \node[right=of mono,
  step node={type\\[-3pt]checking}] (tc) {\faCircleCheck};
  \node[right=of tc,
  step node={overrides\\[-3pt]checking}] (over) {\faCodeBranch};

  \coordinate[right=7mm of over] (out);
  \coordinate (mid) at ($(over)!0.5!(out)$);

  \node[above=2mm of out] (out-ok) {\textcolor{green}{\faCheck}};
  \node[below=2mm of out] (out-no) {\textcolor{red}{\faXmark}};

  \begin{scope}[connect]
    \draw (IR) -- (infer);
    \draw (infer) -- (mono);
    \draw (mono) -- (tc);
    \draw (tc) -- (over);
    \draw (over.east) -- (out-ok);
    \draw (over.east) -- (out-no);
  \end{scope}
  
\end{tikzpicture}
\caption{An overview of \ranger's typechecking workflow.}
\label{fig:type-checking-workflow}
\end{figure}

We implemented \ranger on top of the \licorne experimental programming language. 
The implementation language of both \licorne's compiler and \ranger's type-checker is Scala. 
In the current prototype,
the type checker implementation is approximately 4400 LOC,
whereas the AST-to-IR conversion is about 1300 LOC.
\autoref{fig:type-checking-workflow} shows 
the typechecking workflow, whose individual phases
we described in \autoref{sec:design}. 
It runs on an intermediate
representation that is a more complex version of \autoref{fig:syntax}'s \ircorne.
As seen in the previous sections,
\ranger's bidirectional type checking algorithm
heavily relies on control-flow information
that is more readily available in a bespoke SSA-form IR.
This is why \ranger's implementation
performs type checking at the IR level%
---in contrast to most mainstream programming languages
that perform typechecking on the AST.

\paragraph{Constraint solving.}
The type checker's implementation calls out to the Z3 SMT solver~\cite{z3}
to check type constraints involving ranges and other refinement types:
\begin{enumerate*}[label=\emph{\roman*})]
\item the subtyping relation between refinements involves checking the validity of an implication (rule \rulen{sub-ref} in \autoref{fig:subtyping});
\item monotonicity analysis checks the validity of inequalities involving integer variables updated in a loop;
\item a type simplification mechanism that tries to make error messages 
more readable relies on SMT solving to rewrite range bounds in succinct form;
\item the simple form of abstract interpretation performed by \autoref{fig:typing-rules-expressions}'s 
rule \rulen{arith} also uses SMT solving.
\end{enumerate*}

In general,
the constraints sent to the SMT solver are in the quantifier free fragment
of integer arithmetic and uninterpreted functions theories~\cite{Kroening2016},
which is undecidable (recursively enumerable).
When type annotations only involve range types (not general refinements)
and the corresponding numeric variables are only linearly updated,
then the constraints are in integer \emph{linear} arithmetic,
which is NP complete.
Despite these worst-case bounds,
as confirmed in \autoref{sec:evaluation}'s experiments,
\ranger's typechecking is fast on ``natural'' programs.
In addition, \ranger's implementation reduces the complexity of the constraints
that it reasons about in different ways:
\begin{enumerate*}[label=\emph{\roman*})]
  \item The candidate type inference's dataflow equations (\autoref{tab:candidate-backward-rules}) are not iterated to fixpoint
    but only unrolled once per loop;
    while this may produce unsound inference,
    this is not a problem because the typechecking phase
    validates any candidates before using them.
  \item The type checker uses e-graphs~\cite{e-graphs} to
    transparently propagate typing information to all variables
    that are equal at a certain program point. This leverages
    the flow-sensitive information to avoid
    applying the same typing judgments redundantly.
\end{enumerate*}

\section{Evaluation}
\label{sec:evaluation}

The experimental evaluation targets two research questions:

\begin{description}[leftmargin=22pt,labelsep=2pt,labelindent=0pt,nosep]
\item[RQ1] Does \ranger require a similar number of annotations as standard type systems?
\item[RQ2] How does \ranger compare to similar frameworks in expressiveness, conciseness, soundness, and precision?
\end{description}

\subsection{Experimental Setup}
\label{sec:eval:setup}

\paragraph{Comparable systems.}
\licorne's features are closest to Java, Kotlin, and Scala.
Since the latter has the most advanced type inference,
we compare annotation overhead in Scala and \ranger to investigate RQ1.
For RQ2, we compare \ranger to type systems for similar  languages
that can express ranges and whose implementation is available and usable.
Thus, we select
the Checker Framework~\cite{java_checker_framework} (especially its Index Checker~\cite{java_index_checker}) and Liquid Java~\cite{liquid_java}
as comparable systems.
In contrast, we exclude other refinement type systems mentioned in \autoref{sec:related-work}
because they extend languages with fundamentally different type systems
(e.g., Haskell, TypeScript, Rust)
or their implementations are unmaintained %
(e.g., qualified types for Scala~\cite{qualified_types_scala}).

\paragraph{Subjects.}
We collected a total of 14 example programs in 5 groups%
---listed in \autoref{tab:results}---%
from various sources that showcase different usages of range types. %
We wrote the programs in group \pgroup{rng} ourselves, expressly to demonstrate \ranger's capabilities:
these include
data structures that represent dates and times (\pex{DateTime}),
integer filtering functions (\pex{FilterLess}),
and \autoref{fig:ir-example}'s
and \autoref{fig:motivating-example}
motivating examples (\pex{MaxPos} and \pex{Decoder}).
Group \pgroup{amap}
is a simplified version of the array map implementation
in the Plume graph library~\cite{plume_library},
which we targeted because it
already uses developer-written Checker Framework annotations.
Group \pgroup{sort} is a merge-sort implementation
from~\cite{the_algorithms_java}
annotated for the Checker Framework
by a student of the \emph{Software Analysis} course~\cite{sa-repo};
we revised their annotations to make them as concise as possible.
Subjects in group \pgroup{ic}
are from the Checker Framework manual~\cite[\S~11]{java_checker_framework_manual},
where they demonstrate using the Index Checker
to verify correct index usage in array-manipulating programs.
Programs in group \pgroup{liq} are from
various LiquidJava resources~\cite{liquid_java_examples_repo,liquid_java_implementation_repo,liquid_java_openvsx_page}.
Groups \pgroup{liq} and \pgroup{ic}
collect all meaningful examples we could find in the Index Checker's and
Liquid Java's official documentation and repositories
that feature range refinement types.
Several of the examples deliberately include variants with errors to also validate the soundness of the tools.

\paragraph{Setup.}
\label{parag:setup}
We first encoded each subject in Licorne, Java, and Scala with
standard type annotations;
the translations are behaviorally identical and, as much as possible,
strike a balance
between using each language's idiomatic features
and retaining direct comparability.
In these translations, the input and output parameters
of each \emph{unit}
(a method or function, \autoref{tab:results}'s column \textsc{un})
is annotated with standard types: column \textsc{std} reports the number of annotated parameters.
Second, we refined the basic type annotations with range types
wherever possible according to each framework's capabilities.
Column
\textsc{refin} counts
the number of such annotations in signatures (a proxy for \emph{conciseness}),
and \textsc{constr} the number of underlying
constraints they express (a proxy for \emph{expressiveness}).
For example, the integer range from $0$ to $n$ included
is expressible as \Lic{<<0..n>>} in \ranger
and as \J{@Nonnegative @LessThan("n + 1")} in the Index Checker;
thus, \textsc{constr} is 2 in both (lower and upper bounds),
whereas \textsc{refin} is 1 in \ranger and 2 in the Index Checker.
Third, we added refinement annotations in the body of
the functions as required by each framework in order to 
successfully typecheck the function (or, if it contains a bug,
to locate it);
these \textsc{aux}iliary annotations measure the user's annotation burden.
Sometimes, typechecking only succeeds if we add
explicit casts or other forms of forced assumptions;
we also recorded the number of such \textsc{cas}ts
(except for LiquidJava, which provides no such casting mechanism).
Finally, we ran each framework's typechecking on the examples,
and recorded the outcome for each unit:
\begin{enumerate*}
\item a true positive \textsc{tp} is when a buggy unit fails typechecking,
  and hence the type checker finds the bug;
\item a false positive \textsc{fp} is when a correct unit fails typechecking
  (even if we add explicit casts),
  indicating a completeness failure in the type checker;
\item a true negative \textsc{tn} is when a correct unit typechecks successfully;  
  \item a false negative \textsc{fn} is when a buggy unit typechecks,
  indicating a soundness failure in the type checker.
\end{enumerate*}
To enable a meaningful comparison between different tools
that may report multiple errors differently,
we count a unit as buggy (column \textsc{b}) iff it contains
at least one bug.
While working on these experiments,
the LiquidJava verifier crashed
on several examples (indicated as $\failure$ in \autoref{tab:results});
given these issues, we did not include the two more complex subjects
in LiquidJava's comparison (empty cells in \autoref{tab:results}).\footnote{
  While preparing the examples,
  we contacted the authors of LiquidJava and revised our examples
  incorporating their suggestions on how to use their tool.
}

\begin{table}[!tb]
  \centering
  \scriptsize
  \setlength{\tabcolsep}{1.5pt}
  \renewcommand{\arraystretch}{0.7}
  \begin{tabular}{
    c l
    *{4}{r}
    *{2}{r}
    *{2}{*{3}{r}}
    *{4}{r}
    *{2}{r}
    *{4}{*{3}{r}}
    }
\toprule
    & & & & &
    & \multicolumn{2}{c}{\textsc{loc}}
    & \multicolumn{3}{c}{\textsc{refin}}
    & \multicolumn{3}{c}{\textsc{constr}}
    & \multicolumn{4}{c}{\textsc{aux}}
    & \multicolumn{2}{c}{\textsc{cas}}
    & \multicolumn{3}{c}{\textsc{tp}}
    & \multicolumn{3}{c}{\textsc{fp}}
    & \multicolumn{3}{c}{\textsc{tn}}
    & \multicolumn{3}{c}{\textsc{fn}}
    \\
    \cmidrule(lr){7-8}
    \cmidrule(lr){9-11}
    \cmidrule(lr){12-14}
    \cmidrule(lr){15-18}
    \cmidrule(lr){19-20}
    \cmidrule(lr){21-23}
    \cmidrule(lr){24-26}
    \cmidrule(lr){27-29}
    \cmidrule(lr){30-32}
    \multicolumn{1}{c}{\textsc{\textsc{group}}}
    & \multicolumn{1}{c}{\textsc{prog}}
    & \multicolumn{1}{c}{\textsc{un}}
    & \multicolumn{1}{c}{\textsc{b}} 
    & \multicolumn{1}{c}{\textsc{std}}
    & \multicolumn{1}{c}{\textsc{csr}}
    & \multicolumn{1}{c}{\textsc{R}} & \multicolumn{1}{c}{\textsc{S}}
    & \multicolumn{1}{c}{\textsc{R}} & \multicolumn{1}{c}{\textsc{C}} & \multicolumn{1}{c}{\textsc{L}}
    & \multicolumn{1}{c}{\textsc{R}} & \multicolumn{1}{c}{\textsc{C}} & \multicolumn{1}{c}{\textsc{L}}
    & \multicolumn{1}{c}{\textsc{R}} & \multicolumn{1}{c}{\textsc{C}} & \multicolumn{1}{c}{\textsc{L}} & \multicolumn{1}{c}{\textsc{S}}
    & \multicolumn{1}{c}{\textsc{R}} & \multicolumn{1}{c}{\textsc{C}}
    & \multicolumn{1}{c}{\textsc{R}} & \multicolumn{1}{c}{\textsc{C}} & \multicolumn{1}{c}{\textsc{L}}
    & \multicolumn{1}{c}{\textsc{R}} & \multicolumn{1}{c}{\textsc{C}} & \multicolumn{1}{c}{\textsc{L}}
    & \multicolumn{1}{c}{\textsc{R}} & \multicolumn{1}{c}{\textsc{C}} & \multicolumn{1}{c}{\textsc{L}}
    & \multicolumn{1}{c}{\textsc{R}} & \multicolumn{1}{c}{\textsc{C}} & \multicolumn{1}{c}{\textsc{L}}
    \\
    \midrule
    \multirow{4}{*}{\pgroup{rng}}      & \pex{DateTime}   & 14 &  2 &  30 &  20 &  65 &  61 & 10 & 10 &       10 &  20 &  19 &       20 & 0 & 0 &        0 & 0 & 0 & 0 &  2 &  2 &        0 & 0 & 0 &        0 & 12 & 12 &       12 & 0 & 0 &        2 \\
                                   & \pex{FilterLess} &  4 &  2 &  12 &   4 &  24 &  14 &  4 &  4 &          &   4 &   4 &          & 0 & 2 &          & 0 & 0 & 0 &  2 &  1 &          & 0 & 1 &          &  2 &  1 &          & 0 & 1 &          \\
                                   & \pex{MaxPos}     &  4 &  2 &  14 &  11 &  56 &  56 & 10 &  9 & \failure &  11 &  11 & \failure & 0 & 0 & \failure & 0 & 0 & 0 &  2 &  2 & \failure & 0 & 0 & \failure &  2 &  2 & \failure & 0 & 0 & \failure \\
                                   & \pex{Decoder}    &  8 &  0 &  23 &  17 &  44 &  52 & 12 & 15 & \failure &  17 &  15 & \failure & 0 & 4 & \failure & 4 & 0 & 5 &  0 &  0 & \failure & 0 & 1 & \failure &  8 &  7 & \failure & 0 & 0 & \failure \\
\cmidrule(l){3-32}\multirow{1}{*}{\pgroup{amap}}     & \pex{ArrayMap}   &  8 &  1 &  15 &   9 &  57 &  54 &  6 &  6 &          &   9 &   9 &          & 0 & 0 &          & 0 & 0 & 0 &  1 &  1 &          & 0 & 0 &          &  7 &  7 &          & 0 & 0 &          \\
\cmidrule(l){3-32}\multirow{1}{*}{\pgroup{sort}}     & \pex{Sorting}    &  3 &  1 &  13 &  21 &  37 &  35 &  8 &  6 &        5 &  21 &  21 &       21 & 0 & 1 &        0 & 0 & 1 & 3 &  1 &  1 &        0 & 0 & 0 &        0 &  2 &  2 &        2 & 0 & 0 &        1 \\
\cmidrule(l){3-32}\multirow{4}{*}{\pgroup{ic}}       & \pex{ThirdElem}  &  4 &  2 &   8 &   4 &  12 &  12 &  4 &  4 &        4 &   4 &   4 &        4 & 0 & 0 &        0 & 0 & 0 & 0 &  2 &  2 &        0 & 0 & 0 &        0 &  2 &  2 &        2 & 0 & 0 &        2 \\
                                   & \pex{ArrayLess}  &  2 &  1 &   6 &   2 &  20 &  20 &  2 &  2 &        2 &   2 &   2 &        2 & 0 & 0 &        0 & 0 & 0 & 0 &  1 &  1 &        0 & 0 & 0 &        0 &  1 &  1 &        1 & 0 & 0 &        1 \\
                                   & \pex{RemString}  &  3 &  1 &   9 &   0 &  27 &  20 &  0 &  0 & \failure &   0 &   0 & \failure & 0 & 0 & \failure & 0 & 0 & 0 &  1 &  1 & \failure & 0 & 0 & \failure &  2 &  2 & \failure & 0 & 0 & \failure \\
                                   & \pex{ArrayWrap}  &  6 &  0 &  10 &   8 &  19 &  13 &  5 &  6 & \failure &   8 &   8 & \failure & 0 & 0 & \failure & 1 & 0 & 0 &  0 &  0 & \failure & 0 & 0 & \failure &  6 &  6 & \failure & 0 & 0 & \failure \\
\cmidrule(l){3-32}\multirow{4}{*}{\pgroup{lj}}       & \pex{Fibonacci}  &  2 &  2 &   4 &   6 &  14 &  14 &  6 &  4 &        4 &   6 &   4 &        6 & 0 & 0 &        0 & 0 & 0 & 0 &  2 &  0 &        1 & 0 & 0 &        0 &  0 &  2 &        0 & 0 & 0 &        1 \\
                                   & \pex{Sum}        &  2 &  1 &   4 &   4 &  18 &  18 &  4 &  2 &        2 &   4 &   2 &        4 & 0 & 0 &        0 & 0 & 0 & 0 &  1 &  0 &        1 & 0 & 0 &        0 &  1 &  2 &        1 & 0 & 0 &        0 \\
                                   & \pex{AbsDiv}     &  3 &  1 &   7 &   4 &   9 &   9 &  4 &  3 &        4 &   4 &   2 &        4 & 0 & 0 &        0 & 0 & 0 & 0 &  1 &  0 &        0 & 0 & 0 &        0 &  2 &  3 &        2 & 0 & 0 &        1 \\
                                   & \pex{Car}        &  5 &  1 &   4 &   7 &  22 &  22 &  4 &  4 &        4 &   7 &   7 &        7 & 0 & 0 &        0 & 0 & 0 & 0 &  1 &  1 &        1 & 0 & 0 &        0 &  4 &  4 &        4 & 0 & 0 &        0 \\
\midrule\multicolumn{2}{c}{\textbf{Total}}                 & 68 & 17 & 159 & 117 & 424 & 400 & 79 & 75 &       35 & 117 & 108 &       68 & 0 & 7 &        0 & 5 & 1 & 8 & 17 & 12 &        3 & 0 & 2 &        0 & 51 & 53 &       24 & 0 & 1 &        8 \\

    \bottomrule
\end{tabular}
  \caption{\smaller Evaluation results. Each row refers to a \textsc{prog}ram within a \textsc{group} that implements a number of code \textsc{un}its (methods/functions). %
    For each program, the table reports
    the number \textsc{b} of units with bugs,
    the number \textsc{std} of standard types in signatures
    (typed parameters plus return type),
    the maximal number \textsc{csr} of range-like refinement constraints. %
    The rest of the table reports data for each framework or language: \licorne with \ranger annotations (\textsc{R}), Java with Checker Framework (\textsc{C}) and with LiquidJava (\textsc{L}) annotations, and regular Scala (\textsc{S}).
    These data include:
    the number \textsc{loc} of lines of code of each unit, %
    the number of individual \textsc{refin}ement annotations in signatures
    (a \J{@} annotation for \textsc{C} and \textsc{L}, and
    a range or \Lic{with} for \textsc{R}),
    the number of \textsc{constr}aints these refinements express,
    the number of \textsc{aux}iliary annotations inside the function body,
    the number \textsc{cas} of forced casts (hybrid casts in \textsc{R}, 
    \J{@AssumeAssertion} in \textsc{C}),
    the number of units that are
    true positives (\textsc{tp}), false positives (\textsc{fp}),
    true negatives (\textsc{tn}), and false negatives (\textsc{fn}).
    Dashes $\failure$ denote cases where the tool crashed.}
  \label{tab:results}
\end{table}

\subsection{Experimental Results}

\subsubsection{RQ1.}
To answer RQ1, we compare columns \textsc{aux} in \autoref{tab:results},
which report the number of auxiliary, in-body type annotations required
in the examples.
\ranger (column \textsc{R}) requires no auxiliary annotations,
whereas Scala (column \textsc{S}) requires five.
This clearly indicates that \ranger's expressive features
do not introduce a heavier annotation burden 
 than a standard language's type system.
 In fact, \ranger uses no auxiliary annotations at all,
 whereas Scala does in two examples
 where initializing a generic array or collection
 requires explicit type parameter annotations.
One such example is program \pex{ArrayWrap},
where Scala's behavior is possibly the result of using a \J{ClassTag} object.
The other is program \pex{Decoder},
where there are four occurrences of the common pattern:
instantiate a collection, update it in a loop, and return it.
Here, \ranger's type inference algorithm backpropagates the return type
to the instantiation point, thus removing the need for an explicit annotation.
Interestingly, in these examples Scala can perform \emph{forward} inference
instead, so that one can omit the return type in the signature.
While we may argue that an explicitly typed signature
is more generally useful than an in-body annotation,
it remains clear that using \ranger's type system is no more burdensome or verbose
than using a mainstream language like Scala, while providing additional
static correctness guarantees.

\subsubsection{RQ2.}
Let's compare \ranger (columns \textsc{R} in \autoref{tab:results})
to the Checker Framework and LiquidJava (columns \textsc{C} and \textsc{L})
for expressiveness, soundness, precision, and conciseness.

\paragraph{Expressiveness}
is measured by the fraction $\textsc{constr}/\textsc{csr}$
of constraints expressible in each framework.
According to this metric,
\ranger is
the most expressive (100\% of constraints are expressible)
whereas the Checker Framework can express 92\% of constraints.
LiquidJava also reaches 100\% expressiveness on all the examples
that we could tackle, but its expressiveness is undefined
on the examples where it crashes.
In all, these results demonstrate the higher expressiveness of
systems based on free refinement types---as opposed to
the fixed set of predefined annotations offered by the Checker Framework: 
the Constant Value Checker and Index Checker annotations
can only encode constant variable bounds and
the most common collection access patterns.
For instance, the only lower bounds 
that the Index Checker can express are $-1$, $0$, and $1$.

\paragraph{Conciseness}
is measured as $(\textsc{refin} + \textsc{aux} + \textsc{cas}) / \textsc{constr}$,
namely the average number of all annotations needed to encode
the expressible range constraints: the smaller, the more concise.
\ranger uses 0.68 annotations per constraint,
which is significantly more concise than the Checker Framework's 0.83 annotations
per constraint;
with its 0.51 annotations per constraint,
LiquidJava is the most concise%
---with the important caveat that the practical significance of this result 
is limited by its incompatibility with certain Java features.
As for expressiveness, free refinement type notations
tend to be more succinct than those using a predefined set;
while we did not rigorously evaluate readability,
it is plausible that conciseness is beneficial to it as well.

\paragraph{Soundness}
is measured by the fraction $\textsc{tn} / (\textsc{tn} + \textsc{fn})$
of all correct units that pass typechecking.
\ranger did not incur any false negatives
in our experiments; hence, its soundness is 100\%.
The Checker Framework incurred one \textsc{fn} in \pex{FilterLess},
where it failed to detect an off-by-one bug
when a filter function was applied to a stream;
indeed, we have encountered other examples
that suggest that the Checker Framework's support for the \J{Stream}
class is sometimes limited.
LiquidJava, in contrast,
presented 7 false negatives, which gives a soundness of 76\%.
These failures usually involve type arguments,
which seems to indicate that LiquidJava does not soundly model generics.

\paragraph{Precision}
is measured by the fraction $\textsc{tp} / (\textsc{tp} + \textsc{fp})$
of all found bugs (i.e., buggy units that fail typechecking).
In our experiments,
\ranger achieved 100\% precision,
whereas the Checker Framework was 86\% precise.
LiquidJava also reached 100\% precision on all the examples
that we could tackle;
however, this result should be weighted against
LiquidJava's significant number of soundness failures.
The Checker Framework's precision loss comes down to its
limited constraint reasoning capabilities,
in contrast to \ranger's flexible usage of an SMT solver.
A glaring example of the former's limitation occurs
in \autoref{fig:motivating-example}'s example,
where it cannot verify without a forced assumption
that the second argument in the call to \J{clamp}
in \J{decode} is nonnegative:
its solver cannot check the equality $\J{xLen - 1 + 1} = \J{xLen}$.
Another example
is when instantiating two arrays \J{a}, \J{b} with the same length \J{s}:
the Checker Framework only recognizes this fact if \J{b}
is initialized as \J{new int[a.length]},
but not if as \J{new int[s]};
\ranger supports both patterns as equivalent.

\subsubsection{Performance.}
While a thorough benchmarking of \ranger's performance
is outside the scope of this paper,
we collected basic performance data to ascertain
that its typechecking algorithm is reasonably fast in practice.
In our experiments,
\ranger typechecked all the 14 examples in \autoref{tab:results}
in 16s;
by comparison,
the Scala~3 compiler took 34s to typecheck
the same examples translated to Scala
(note that we only measured the \emph{typechecking}
time, not the whole compilation time).
These results indicate that
\ranger's typechecking process is practically usable,
and that its underlying program analyses are
generally efficient.

\section{Discussion and Conclusions}
\label{sec:discussion}

\ranger is an approach to range refinement types that
focuses on ease of use and light annotation overhead
rather than maximal expressiveness;
in other words, it tries to extend the
capabilities of standard type systems while
retaining their ergonomics.
This led to a type system that supports
features like flow sensitivity with a custom
backward inference algorithm, which significantly lowers
the user annotation burden while naturally capturing
common programming idioms.
In our experiments, \ranger
used annotations in measure comparable to
a standard language's type system;
and provided more expressivity, conciseness, or capabilities
than comparable range refinement type systems.
These results also showcase the advantages
of weaving the design of a refinement type systems
organically into a language's idioms---as opposed
to retrofitting it ex post.
In future work, we plan to generalize our approach to
other kinds of refinement types %
and try to implement it for existing programming languages.

\clearpage
\bibliographystyle{splncs04}
\bibliography{range-types}

@inproceedings{Callauetal2014,
  author    = {Calla\'{u}, Oscar and Robbes, Romain and Tanter, \'{E}ric and R\"{o}thlisberger, David and Bergel, Alexandre},
  title     = {On the Use of Type Predicates in Object-Oriented Software: The Case of {Smalltalk}},
  booktitle = {Proc. 10th ACM Symposium on Dynamic Languages},
  year      = {2014},
  pages     = {135--146},
  publisher = {ACM},
  doi       = {10.1145/2661088.2661091}
}

@book{Pierce2002,
  author    = {Benjamin C. Pierce},
  title     = {Types and Programming Languages},
  publisher = {The {MIT} Press},
  year      = {2002},
  isbn      = {978-0-262-16209-8}
}

@inproceedings{Bruijn70,
  author    = {Nicolaas Govert de Bruijn},
  title     = {The Mathematical Language {AUTOMATH}, Its Usage, and Some of Its Extensions},
  booktitle = {Symposium on Automatic Demonstration},
  series    = {Lecture Notes in Mathematics},
  volume    = {125},
  pages     = {29--61},
  publisher = {Springer},
  year      = {1970},
  doi       = {10.1007/BFb0060623}
}

@book{intuitionistic,
  author    = {Per Martin{-}L{\"{o}}f},
  title     = {Intuitionistic type theory},
  series    = {Studies in proof theory},
  volume    = {1},
  publisher = {Bibliopolis},
  year      = {1984},
  isbn      = {978-88-7088-105-9}
}

@article{coq,
  author    = {Thierry Coquand and G{\'{e}}rard P. Huet},
  title     = {The Calculus of Constructions},
  journal   = {Inf. Comput.},
  volume    = {76},
  number    = {2/3},
  pages     = {95--120},
  year      = {1988},
  doi       = {10.1016/0890-5401(88)90005-3}
}

@article{idris,
  author    = {Edwin Brady},
  title     = {{Idris}, a general-purpose dependently typed programming language: Design and implementation},
  journal   = {J. Funct. Program.},
  volume    = {23},
  number    = {5},
  pages     = {552--593},
  year      = {2013},
  doi       = {10.1017/S095679681300018X}
}

@inproceedings{CousotC77,
  author       = {Patrick Cousot and
                  Radhia Cousot},
  title        = {Abstract Interpretation: {A} Unified Lattice Model for Static Analysis
                  of Programs by Construction or Approximation of Fixpoints},
  booktitle    = {Conference Record of the Fourth {ACM} Symposium on Principles of Programming
                  Languages, Los Angeles, California, USA, January 1977},
  pages        = {238--252},
  publisher    = {{ACM}},
  year         = {1977},
  OPTurl          = {https://doi.org/10.1145/512950.512973},
  doi          = {10.1145/512950.512973},
  bibsource    = {dblp computer science bibliography, https://dblp.org}
}

@article{e-graphs,
  author       = {Max Willsey and
                  Chandrakana Nandi and
                  Yisu Remy Wang and
                  Oliver Flatt and
                  Zachary Tatlock and
                  Pavel Panchekha},
  title        = {{egg}: Fast and extensible equality saturation},
  journal      = {Proc. {ACM} Program. Lang.},
  volume       = {5},
  number       = {{POPL}},
  pages        = {1--29},
  year         = {2021},
  url          = {https://doi.org/10.1145/3434304},
  doi          = {10.1145/3434304},
  bibsource    = {dblp computer science bibliography, https://dblp.org}
}

@book{Kroening2016,
  author    = {Kroening, Daniel and Strichman, Ofer},
  title     = {Decision Procedures: An Algorithmic Point of View},
  publisher = {Springer},
  year      = {2016},
  edition   = {2nd},
  doi       = {10.1007/978-3-662-50497-0}
}

@article{DBLP:journals/sigplan/Appel88,
  author       = {Andrew W. Appel},
  title        = {{SSA} is Functional Programming},
  journal      = {{ACM} {SIGPLAN} Notices},
  volume       = {33},
  number       = {4},
  pages        = {17--20},
  year         = {1998},
  doi          = {10.1145/278283.278285},
  bibsource    = {dblp computer science bibliography, https://dblp.org}
}

@inproceedings{JonesWEV16,
  author       = {Simon Peyton Jones and
                  Stephanie Weirich and
                  Richard A. Eisenberg and
                  Dimitrios Vytiniotis},
  editor       = {Sam Lindley and
                  Conor McBride and
                  Philip W. Trinder and
                  Donald Sannella},
  title        = {A Reflection on Types},
  booktitle    = {A List of Successes That Can Change the World - Essays Dedicated to
                  Philip Wadler on the Occasion of His 60th Birthday},
  series       = {Lecture Notes in Computer Science},
  pages        = {292--317},
  publisher    = {Springer},
  year         = {2016},
  OPTurl          = {https://doi.org/10.1007/978-3-319-30936-1\_16},
  doi          = {10.1007/978-3-319-30936-1\_16},
  bibsource    = {dblp computer science bibliography, https://dblp.org}
}

@inproceedings{DBLP:conf/icfp/Tobin-HochstadtF10,
  author    = {Sam Tobin{-}Hochstadt and Matthias Felleisen},
  title     = {Logical types for untyped languages},
  booktitle = {Proc. 15th {ACM} {SIGPLAN} International Conference on Functional Programming, {ICFP} 2010, Baltimore, Maryland, USA, September 27-29, 2010},
  pages     = {117--128},
  publisher = {{ACM}},
  year      = {2010},
  doi       = {10.1145/1863543.1863561}
}

@inproceedings{DBLP:conf/popl/Flanagan06,
  author    = {Cormac Flanagan},
  title     = {Hybrid type checking},
  booktitle = {Proc. 33rd {ACM} {SIGPLAN-SIGACT} Symposium on Principles of Programming Languages, {POPL} 2006, Charleston, South Carolina, USA, January 11-13, 2006},
  pages     = {245--256},
  publisher = {{ACM}},
  year      = {2006},
  doi       = {10.1145/1111037.1111059}
}

@article{liquid_types_ml,
  author    = {Tim Freeman and Frank Pfenning},
  title     = {Refinement types for {ML}},
  journal   = {SIGPLAN Not.},
  volume    = {26},
  number    = {6},
  month     = {jun},
  year      = {1991},
  pages     = {268--277},
  doi       = {10.1145/113446.113468}
}

@inproceedings{liquid_types_ocaml,
  author    = {Patrick M. Rondon and Ming Kawaguci and Ranjit Jhala},
  title     = {Liquid types},
  booktitle = {Proc. 29th ACM SIGPLAN Conference on Programming Language Design and Implementation (PLDI '08)},
  publisher = {Association for Computing Machinery},
  address   = {New York, NY, USA},
  year      = {2008},
  pages     = {159--169},
  doi       = {10.1145/1375581.1375602}
}

@inproceedings{liquid_haskell,
  author    = {Niki Vazou and Eric L. Seidel and Ranjit Jhala and Dimitrios Vytiniotis and Simon Peyton-Jones},
  title     = {Refinement types for {Haskell}},
  booktitle = {Proc. 19th ACM SIGPLAN International Conference on Functional Programming (ICFP '14)},
  publisher = {Association for Computing Machinery},
  address   = {New York, NY, USA},
  year      = {2014},
  pages     = {269--282},
  doi       = {10.1145/2628136.2628161}
}

@inproceedings{aeon,
  title={The usability argument for refinement typed genetic programming},
  author={Fonseca, Alcides and Santos, Paulo and Silva, Sara},
  booktitle={International Conference on Parallel Problem Solving from Nature},
  pages={18--32},
  year={2020},
  organization={Springer}
}

@article{flux,
  author    = {Nico Lehmann and Adam T. Geller and Niki Vazou and Ranjit Jhala},
  title     = {Flux: Liquid Types for {Rust}},
  journal   = {Proc. ACM Program. Lang.},
  volume    = {7},
  number    = {PLDI},
  month     = {jun},
  year      = {2023},
  pages     = {169:1--169:25},
  doi       = {10.1145/3591283}
}

@inproceedings{refined_typescript,
  author    = {Panagiotis Vekris and Benjamin Cosman and Ranjit Jhala},
  title     = {Refinement types for {TypeScript}},
  booktitle = {Proc. 37th ACM SIGPLAN Conference on Programming Language Design and Implementation (PLDI '16)},
  publisher = {Association for Computing Machinery},
  address   = {New York, NY, USA},
  year      = {2016},
  pages     = {310--325},
  doi       = {10.1145/2908080.2908110}
}

@inproceedings{qualified_types_scala,
  author    = {Georg Stefan Schmid and Viktor Kuncak},
  title     = {{SMT}-based checking of predicate-qualified types for {Scala}},
  booktitle = {Proc. 2016 7th ACM SIGPLAN Symposium on Scala (SCALA 2016)},
  publisher = {Association for Computing Machinery},
  address   = {New York, NY, USA},
  year      = {2016},
  pages     = {31--40},
  doi       = {10.1145/2998392.2998398}
}

@mastersthesis{valentin_msc_thesis,
  title        = {Implementation of a gradually compartmentalized programming language},
  author       = {Valentin Aebi},
  year         = 2025,
  month        = {January},
  address      = {Lausanne, Switzerland},
  note         = {Available at \url{https://valentinaebi.github.io/resources/ms-thesis.pdf}},
  school       = {EPFL},
  type         = {Master's thesis}
}

@inproceedings{java_index_checker,
  author    = {Martin Kellogg and Vlastimil Dort and Suzanne Millstein and Michael D. Ernst},
  title     = {Lightweight verification of array indexing},
  booktitle = {Proc. 27th ACM SIGSOFT International Symposium on Software Testing and Analysis (ISSTA 2018)},
  publisher = {Association for Computing Machinery},
  address   = {New York, NY, USA},
  year      = {2018},
  pages     = {3--14},
  doi       = {10.1145/3213846.3213849}
}

@inproceedings{java_checker_framework,
  author    = {Matthew M. Papi and Mahmood Ali and Telmo Luis Correa and Jeff H. Perkins and Michael D. Ernst},
  title     = {Practical pluggable types for {Java}},
  booktitle = {Proc. 2008 International Symposium on Software Testing and Analysis (ISSTA '08)},
  publisher = {Association for Computing Machinery},
  address   = {New York, NY, USA},
  year      = {2008},
  pages     = {201--212},
  doi       = {10.1145/1390630.1390656}
}

@misc{java_checker_framework_manual,
  key = {CheckerFramework},
  title        = {The {Checker Framework} Manual: Custom pluggable types for {Java}},
  howpublished = {\url{https://checkerframework.org/manual/}},
  note         = {Last accessed 2026/06/27}
}

@misc{plume_library,
  key = {Plume},
  title        = {{P}lume Documentation},
  howpublished = {\url{https://github.com/plume-lib/plume-util}},
  note         = {Last accessed 2026/06/21}
}

@misc{the_algorithms_java,
    key = {TheAlgorithms},
    title        = {{T}he {A}lgorithms - {J}ava},
    howpublished = {\url{https://github.com/TheAlgorithms/Java}},
    note = {Last accessed 2026/06/21}
}

@misc{sa-repo,
  author = {Carlo A. Furia},
  title        = {Software Analysis course at {USI}: course material},
  howpublished = {\url{https://github.com/bugcounting/software-analysis/}},
}

@inproceedings{liquid_java,
  author    = {Catarina Gamboa and Paulo Canelas and Christopher Timperley and Alcides Fonseca},
  title     = {Usability-Oriented Design of Liquid Types for {Java}},
  booktitle = {Proc. 45th International Conference on Software Engineering (ICSE '23)},
  publisher = {IEEE Press},
  year      = {2023},
  pages     = {1520--1532},
  doi       = {10.1109/ICSE48619.2023.00132}
}

@inproceedings{csolve,
  author    = {P. Rondon and A. Bakst and M. Kawaguchi and R. Jhala},
  title     = {{Csolve}: Verifying {C} with liquid types},
  booktitle = {International Conference on Computer Aided Verification},
  month     = {jul},
  year      = {2012},
  pages     = {744--750},
  publisher = {Springer Berlin Heidelberg},
  address   = {Berlin, Heidelberg}
}

@inproceedings{refinement_types_ruby,
  author    = {M. Kazerounian and N. Vazou and A. Bourgerie and J. S. Foster and E. Torlak},
  title     = {Refinement types for {Ruby}},
  booktitle = {International Conference on Verification, Model Checking, and Abstract Interpretation},
  month     = {dec},
  year      = {2017},
  pages     = {269--290},
  publisher = {Springer International Publishing},
  address   = {Cham}
}

@article{bidirectional_typing,
  author  = {Jana Dunfield and Neel Krishnaswami},
  title   = {Bidirectional Typing},
  journal = {ACM Comput. Surv.},
  volume  = {54},
  number  = {5},
  month   = {jun},
  year    = {2021},
  pages   = {98:1--98:38},
  doi     = {10.1145/3450952}
}

@inproceedings{z3,
  author    = {L. de Moura and N. Bjørner},
  title     = {{Z3}: An Efficient {SMT} Solver},
  booktitle = {Tools and Algorithms for the Construction and Analysis of Systems. TACAS 2008},
  series    = {Lecture Notes in Computer Science},
  volume    = {4963},
  publisher = {Springer},
  address   = {Berlin, Heidelberg},
  year      = {2008},
  doi       = {10.1007/978-3-540-78800-3_24}
}

@misc{liquid_java_implementation_repo,
  key = {LiquidJava},
  title        = {{L}iquidJava - {E}xtending {J}ava with {L}iquid {T}ypes, {GitHub} repository},
  howpublished = {\url{https://github.com/liquid-java/liquidjava}},
  note         = {Last accessed 2026/06/21}
}

@misc{liquid_java_examples_repo,
  key = {LiquidJava-examples},
  title        = {{LiquidJava} Examples, {GitHub} repository},
  howpublished = {\url{https://github.com/liquid-java/liquidjava-examples}},
  note         = {Last accessed 2026/05/20}
}

@misc{liquid_java_openvsx_page,
  key = {LiquidJava-OpenVSX},
  title        = {{Open VSX page of the {L}iquid{J}ava {V}isual {S}tudio {C}ode extension}},
  howpublished = {\url{https://open-vsx.org/extension/AlcidesFonseca/liquid-java}},
  note         = {Last accessed 2026/06/21}
}

\end{document}